%% file: iui26-20.tex
\documentclass[sigconf]{acmart}
\AtBeginDocument{%
  \providecommand\BibTeX{{%
    \normalfont B\kern-0.5em{\scshape i\kern-0.25em b}\kern-0.8em\TeX}}}

\copyrightyear{2026}
\acmYear{2026}
\setcopyright{cc}
\setcctype{by-nc-nd}
\acmConference[IUI '26]{31st International Conference on Intelligent User Interfaces}{March 23--26, 2026}{Paphos, Cyprus}
\acmBooktitle{31st International Conference on Intelligent User Interfaces (IUI '26), March 23--26, 2026, Paphos, Cyprus}
\acmPrice{}
\acmDOI{10.1145/3742413.3789079}
\acmISBN{979-8-4007-1984-4/2026/03}

\usepackage{colortbl}

\usepackage{graphicx}

\usepackage{balance}
\usepackage{pgffor}
\usepackage{booktabs}       
\usepackage{amsmath}        
\usepackage{hyperref}       
\usepackage{tikz}

\usepackage{amssymb}
\usepackage{xkeyval}
\usepackage{listings}
\usepackage{subcaption}
\usepackage{multirow}
\usepackage{booktabs}

\newcommand{\pvalue}[1]{%
  \ifdim#1pt>0.05pt%
    \textit{p} $={#1}$%
  \else%
    \ifdim#1pt<0.001pt%
        \textit{\textbf{p}} $\mathbf{<.001}$%
    \else
        \textit{\textbf{p}} $\mathbf{={#1}}$%
    \fi
  \fi
}

  {\list{}{\leftmargin=0.05in\itshape}\item[]}%
  {\endlist}

\long\def\comment#1{}

\newcommand{\code}[1]{%
  \colorbox{black!10}{\texttt{#1}}%
}

\newcommand{\nestedcode}[1]{%
  \colorbox{black!20}{\texttt{#1}}%
}

\newcommand{\sysname}{\textsc{Arc}}

\begin{document}

\title[Designing for Strategic Exploration and Responsible AI in SLR]{From Toil to Thought: Designing for Strategic Exploration and Responsible AI in Systematic Literature Reviews}

\author{Runlong Ye}
\affiliation{
  \institution{Computer Science, \\University of Toronto}
  \city{Toronto}
  \state{Ontario}
  \country{Canada}
}
\orcid{0000-0003-1064-2333}
\email{harryye@cs.toronto.edu}

\author{Naaz Sibia}
\affiliation{
  \institution{Computer Science, \\University of Toronto}
  \city{Toronto}
  \state{Ontario}
  \country{Canada}
}
\orcid{0000-0001-7628-7077}
\email{naaz.sibia@utoronto.ca}

\author{Angela Zavaleta Bernuy}
\affiliation{
  \institution{Computing and Software, \\McMaster University}
  \city{Hamilton}
  \state{Ontario}
  \country{Canada}
}
\orcid{0000-0002-1228-5774}
\email{zavaleta@mcmaster.ca}

\author{Tingting Zhu}
\affiliation{
  \institution{Geography, Geomatics and Environment, \\University of Toronto Mississauga}
  \city{Mississauga}
  \state{Ontario}
  \country{Canada}
}
\orcid{0000-0002-3508-1379}
\email{tingting.zhu@utoronto.ca}

\author{Carolina Nobre}
\affiliation{
  \institution{Computer Science, \\University of Toronto}
  \city{Toronto}
  \state{Ontario}
  \country{Canada}
}
\orcid{0000-0002-2892-0509}
\email{cnobre@cs.toronto.edu}

\author{Viktoria Pammer-Schindler}
\affiliation{
  \institution{Graz University of Technology \& Know-Center GmbH}
  \city{Graz}
  \country{Austria}
}
\orcid{0000-0001-7061-8947}
\email{viktoria.pammer-schindler@tugraz.at}

\author{Michael Liut}
\affiliation{
  \institution{Mathematical and Computational Sciences, \\University of Toronto Mississauga}
  \city{Mississauga}
  \state{Ontario}
  \country{Canada}
}
\orcid{0000-0003-2965-5302}
\email{michael.liut@utoronto.ca}

\renewcommand{\shortauthors}{R. Ye, N. Sibia, A. Zavaleta Bernuy, T. Zhu, C. Nobre, \& M. Liut}

\begin{abstract}

Systematic Literature Reviews (SLRs) are fundamental to scientific progress, yet the process is hindered by a fragmented tool ecosystem that imposes a high cognitive load. This friction suppresses the iterative, exploratory nature of scholarly work. To investigate these challenges, we conducted an exploratory design study with 20 experienced researchers. This study identified key friction points: 1) the high cognitive load of managing iterative query refinement across multiple databases, 2) the overwhelming scale and pace of publication of modern literature, and 3) the tension between automation and scholarly agency.

Informed by these findings, we developed \sysname{}, a design probe that operationalizes solutions for multi-database integration, transparent iterative search, and verifiable AI-assisted screening. A comparative user study with 8 researchers suggests that an integrated environment facilitates a transition in scholarly work, moving researchers from managing administrative overhead to engaging in strategic exploration. By utilizing external representations to scaffold strategic exploration and transparent AI reasoning, our system supports verifiable judgment, aiming to augment expert contributions from initial creation through long-term maintenance of knowledge synthesis.
\end{abstract}



\vspace{-1em}
\begin{teaserfigure}
  \includegraphics[width=.55\textwidth]{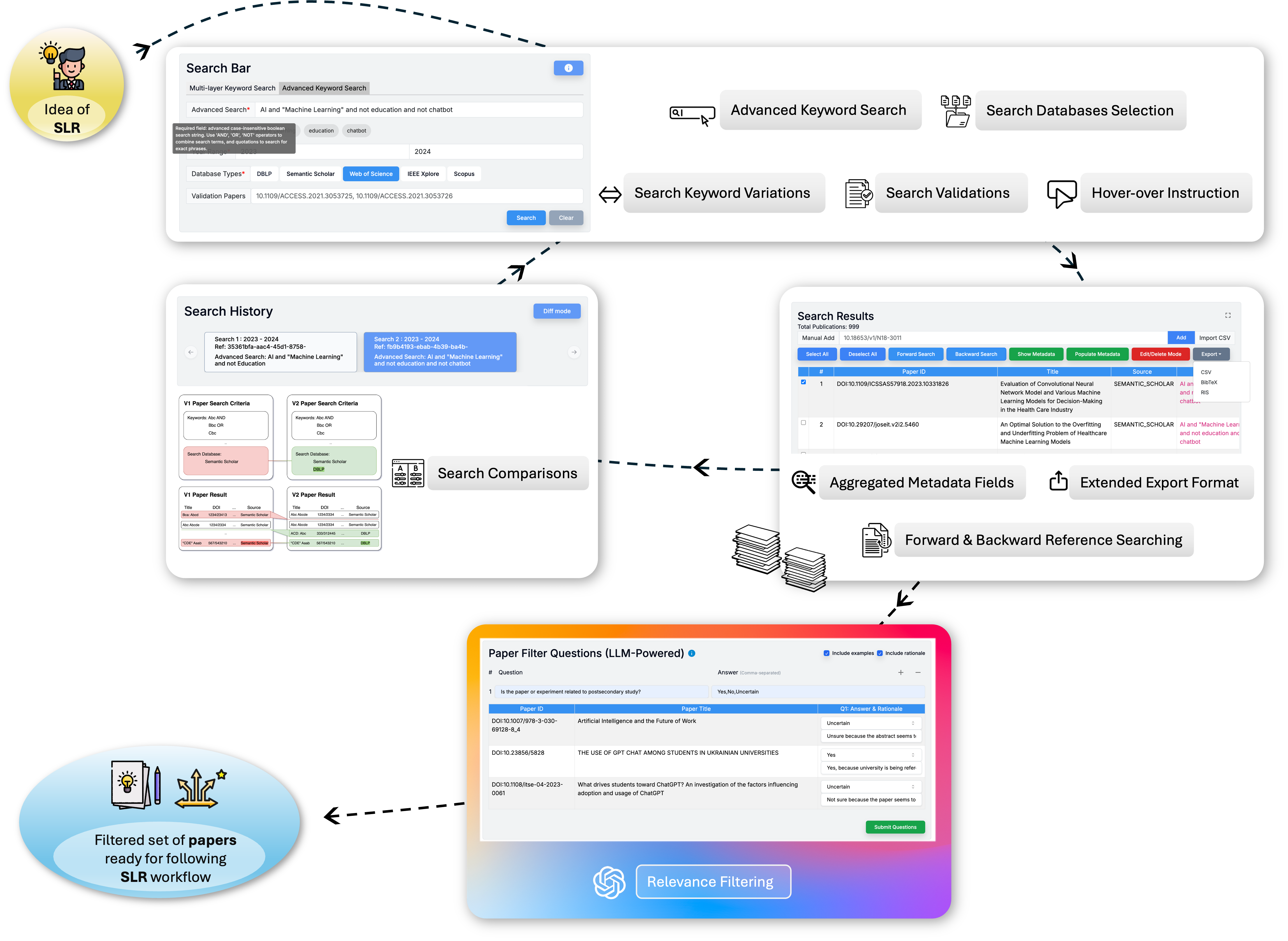}
  \centering
  \vspace{-.5em}
  \caption{\sysname{} is an integrated, human-in-the-loop probe designed to address key challenges in systematic literature review (SLR) creation. The workflow is supported by four core features: (1) a unified search interface to reduce the friction of querying multiple databases; (2) transparent search comparison to support strategic, iterative refinement; (3) integrated reference searching to streamline discovery; and (4) a verifiable, human-in-the-loop AI filtering module for strategic oversight.}
  \Description{This figure illustrates the integrated workflow of the Arc system. The process begins with the "Idea of an SLR," which leads into a unified "Search Bar" that supports multi-layer keyword searches. From this central point, the workflow branches into several features: "Search History" allows users to compare search versions, while "Search Results" displays aggregated metadata. The system also connects to a "Forward and Backward Reference Searching" module. Finally, an "LLM-Powered" section helps users filter papers based on specific questions, leading to a "Relevance Filtering" step that produces a final filtered set of papers ready for the systematic literature review workflow.}
  \label{fig:teaser}
\end{teaserfigure}

\maketitle


\section{Introduction}
\label{intro}
Systematic Literature Reviews (SLRs) provide rigorous, transparent methods for mapping research fields \cite{kitchenham2007guidelines}. However, the process is inherently iterative, requiring researchers to continually refine questions and strategies as they engage with the literature \cite{Bates1989TheDO}. The current ecosystem of academic tools often fails to support this non-linear nature. Researchers face a fragmented landscape of ``point solutions'' that handle discrete tasks, such as citation chasing \cite{citationchaser} or abstract screening \cite{chai2021research}, but lack integration. This forces researchers to manually orchestrate a disjointed workflow across inconsistent database syntaxes and spreadsheets. The resulting cognitive load not only makes the process laborious but discourages the exploration and refinement essential for high-quality reviews \cite{10.1186/s13643-017-0644-y, 10.1145/3706598.3714047}.

To address these challenges, we employed an exploratory design methodology. We first conducted a study with 20 experienced computing researchers to uncover specific bottlenecks in SLR creation. Informed by these findings, we developed \sysname{} (the Automated Review Companion), a high-fidelity design probe built to facilitate a cohesive, intelligent workflow and investigate its impact on research practices.

\sysname{} operationalizes solutions to these challenges by unifying the fractured SLR workflow into a single environment. To streamline discovery, \sysname{} provides an integrated multi-database search that simultaneously queries multiple repositories and automates reference snowballing. This process is supported by an interface designed for strategic iteration, allowing researchers to visually compare search versions and understand the precise impact of query changes. Finally, to mitigate information overload, the system features a transparent, verifiable AI-assisted screening module. This human-in-the-loop component leverages large language models (LLMs) to suggest relevance tags, while strictly upholding researcher agency by providing explicit justifications and requiring user verification.

Driven by the design goals identified in our formative study, this work evaluates the impact of specific interaction mechanisms on the SLR workflow. We investigate:

\begin{itemize}
\item[\textbf{RQ1:}] How can interface design scaffolding support strategic iteration in multi-database searching?
\item[\textbf{RQ2:}] How does automated snowballing integration impact the temporal and cognitive costs of literature discovery?
\item[\textbf{RQ3:}] What are the effects of rationale-based AI suggestions on user trust and verification behavior in systematic screening?
\end{itemize}

To evaluate these questions, we conducted a comparative user study ($N = 8$) comparing \sysname{} against a conventional baseline. Our results indicate that \sysname{} shifts the researcher's focus from manual data maintenance to strategic oversight, transforming query refinement into a systematic exploration. In the discussion, we argue that these findings point toward a new design agenda for next-generation academic tools, one that prioritizes sustainable, interdisciplinary knowledge infrastructures. We conclude with a call to action for the open data ecosystem required to support this vision.

This paper makes three primary contributions: (1) \textit{qualitative insights} from an exploratory study identifying cognitive load and agency tensions as critical SLR bottlenecks; (2) \sysname{}, an \textit{open-source} design probe\footnote{Project repository: \href{https://github.com/CORE-Research-Lab/automated-review-companion}{https://github.com/CORE-Research-Lab/automated-review-companion}.} that operationalizes design principles for responsible review; and (3) \textit{empirical evidence} demonstrating how an integrated, human-in-the-loop environment fosters confident, strategic literature exploration.\enlargethispage*{12pt}

\section{Related Work}

Our work builds upon established methodologies in information science, interactive system design in human-computer interaction (HCI), and recent advances in AI-powered tools. This section first outlines the procedural and cognitive demands of Systematic Literature Reviews (SLRs) to establish the core challenges, then surveys the state-of-the-art of existing tools, revealing a persistent pattern of fragmented support that motivates our design-led investigation into a more integrated, human-centered workflow.

\subsection{Methodology of Systematic Literature Reviews}
Systematic literature reviews (SLRs) are fundamental to scientific progress, serving to synthesize existing knowledge, identify gaps, and provide a foundation for new research \cite{WebsterWatson2002}. Reviews vary widely in their approach, from broad narrative overviews to highly structured systematic reviews \cite{grant2009typology, snyder2019literature}. Our work focuses on SLRs, which are distinguished by their emphasis on rigor, transparency, and reproducibility, with specific methodological adaptations established for fields like computing \cite{kitchenham2007guidelines}.

SLRs employ a formalized, multi-stage process governed by comprehensive guidelines like the PRISMA statement and the Cochrane Handbook \cite{Pagen71, higgins2008cochrane}. However, despite these well-known standards, meta-research reveals a persistent gap between methodological ideals and practice. The production of systematic reviews has reached such a scale that many are redundant or methodologically flawed, and reproducible research practices remain underused \cite{ioannidis2016mass, page2018reproducible, page2018flaws}. Within HCI, umbrella reviews of SLRs have highlighted similar issues with reporting quality and methodological consistency \cite{10.1145/3685266}, and more recent empirical studies have documented the significant practical challenges researchers face in keeping living reviews up-to-date \cite{10.1145/3706598.3714047}. The search process is a notorious bottleneck, researchers must navigate inconsistent syntaxes across databases and meticulously document their strategies according to specialized standards like PRISMA and the PRESS guidelines \cite{ rethlefsen2021prisma, mcgowan2016press}. This procedural complexity is compounded by the high cognitive load inherent in such tasks, which align with foundational theories of information seeking that frame the process not as a simple query-and-retrieval action, but as an iterative process of exploration and sensemaking \cite{Bates1989TheDO, 10.1145/1121949.1121979, pirolli1999information}.\enlargethispage*{12pt}

\subsection{The Landscape of Systematic Literature Review Support}
A variety of tools have been designed to alleviate the burden of specific SLR tasks, but they typically target discrete steps in isolation. The need for better tooling is well-documented; systematic reviews of the SLR process itself have identified the time-consuming nature of the workflow and the difficulties of comprehensive database searching as major challenges \cite{KITCHENHAM20132049s}. In response, the tool landscape has evolved into two primary categories: (1) those that automate procedural steps, and (2) those designed to support higher-level cognitive sensemaking.

\subsubsection{Tools for Automating Procedural Steps:} A significant body of work targets the automation of laborious SLR stages. To address disparate search syntaxes, specialized tools like the Polyglot Search Translator assist researchers in adapting a single query across multiple platforms \cite{clark2020improving}. Other tools target supplementary search methods. For instance, Paperfetcher automates ``handsearching'' specific journals, while Citationchaser focuses on automating citation chasing \cite{paperfetcher, citationchaser}. For the initial collection of literature, tools such as SmartLitReview have been created to automate retrieval from specific, high-volume sources \cite{SmartLitReview}. Once a corpus is collected, the screening phase presents another bottleneck. A range of platforms now uses machine learning to assist with this task, from widely-used web applications like Rayyan and ResearchScreener to open-source frameworks designed to accelerate evidence synthesis \cite{rayyan, chai2021research, van2021open, 10.1186/s13643-019-1074-9}. While effective, these systems are ``point-solutions'' that address discrete stages of the review, leaving the integration and management of the overall workflow to the researcher.

\subsubsection{HCI Systems for Supporting Knowledge Sensemaking:} Recognizing the diversity of review types and the cognitive challenges facing researchers \cite{10.1145/3544548.3581332}, the HCI community has a rich history of developing systems to support the broader work of scholarly sensemaking. Foundational work in this area established the value of interfaces for collaborative search and explored the challenges of achieving a shared understanding \cite{10.1145/1294211.1294215, 10.1145/1518701.1518974, 10.1145/1718918.1718976}. This user-centered tradition continues in modern systems; projects like LitSonar \cite{LitSonar} and LitSense \cite{LitSense}, for example, have conducted formative studies of researcher practices to derive design principles and create tools that better align with existing workflows. Recent systems have introduced novel interactions for discovery, such as organizing literature into conceptual threads \cite{kang2022threddy}, using author-centric views to explore research communities \cite{10.1145/3544548.3581371}, and scaffolding mixed-initiative knowledge synthesis \cite{synergi, 10.1145/3707640.3731913}. Other systems provide valuable ``point-solutions'' for even more granular tasks, such as augmenting the reading of a single paper with personalized context or intelligent skimming support \cite{10.1145/3544548.3580847, 10.1145/3581641.3584034}, scaffolding the synthesis of existing related work sections \cite{10.1145/3544548.3580841}, and enriching paper alerts \cite{10.1145/3613904.3642196}. While these HCI systems are innovative, they are not designed to support the full protocol-driven SLR workflow. This landscape of powerful yet disconnected tools underscores the need for a holistic system that unifies these disparate tasks into a single, coherent workflow.

\subsubsection{Emerging Generative Research Agents} 
Recently, the landscape has expanded to include agentic and AI-enabled systems designed for literature discovery, triage, and evidence appraisal. Tools like Undermind.ai\footnote{https://www.undermind.ai/}, Elicit\footnote{https://elicit.com/}, and Consensus\footnote{https://consensus.app/} leverage LLMs to support natural language search and synthesis, with Elicit additionally offering structured workflows for screening and data extraction. Complementary discovery tools, such as Research Rabbit\footnote{https://www.researchrabbit.ai/}, emphasize citation-network exploration and recommendation from seed papers. Other services, like scite\footnote{https://scite.ai/}, provide ``smart citations'' that surface how a work is cited (e.g., supporting/contrasting/mentioning) to help assess the evidentiary status of claims. General-purpose AI agents such as OpenAI’s ``Deep Research''\footnote{https://openai.com/index/introducing-deep-research/} utilize web-browsing capabilities to aggregate information. However, despite their effectiveness for exploratory scoping or rapid Q\&A, these systems generally lack the structural rigor required for SLRs. They can obscure key aspects of the search strategy (complicating standard practices like PRISMA reporting), struggle with precise Boolean logic, and prioritize textual synthesis over systematic retrieval. Unlike these opaque, fully automated paradigms, our work focuses on scaffolding researcher agency, using AI not to bypass the systematic workflow, but to provide transparency and reproducibility, which are foundational for formal systematic reviews.

In summary, the HCI and broader software engineering communities have a rich history of developing ``point solutions'' for discrete scholarly tasks. While recent empirical work has begun to diagnose the challenges of review maintenance, a critical gap remains. A design-focused investigation is needed into the holistic process of SLR creation. The current literature has well-documented the challenges of fragmentation and cognitive load. However, it has not yet explored, through the construction and evaluation of an integrated system, on how to resolve these issues, nor how such an intervention might reshape researcher strategy. Our work addresses this gap directly.

\section{Exploratory Design Study}
To understand the current practices, challenges, and needs of real-world researchers conducting Systematic Literature Reviews (SLRs), we conducted an exploratory design study with 20 experienced computing researchers. Our goal was to identify critical gaps in existing workflows and tool support, thereby deriving a set of principled design goals to guide the development of a probe that supports rigorous, efficient, and transparent literature analysis.

While the challenges of the systematic review lifecycle are well-documented, from procedural bottlenecks in the initial creation process \cite{KITCHENHAM20132049s} to the difficulties of long-term maintenance \cite{10.1145/3706598.3714047}, our study was designed to elicit deeper, contextual insights into the lived experiences and workflow friction faced by researchers during the creation phase, aiming to understand the why behind these established needs.

\subsection{Study Design and Rationale}
\label{sec:formative-design}
We conducted semi-structured interviews with 20 experienced computing researchers (R1–R20) from diverse subfields, including Machine Learning, HCI, and Databases (see Table \ref{tab:interview-metadata} for researcher demographics). To ground our interviews in concrete functionalities and move beyond abstract discussions of workflow, we developed a series of concept mock-ups illustrating an integrated SLR tool. These were presented to participants as a tangible artifact to provoke discussion and elicit specific feedback. The mock-ups visually materialized key stages of a potential future workflow, including: (1) complex search formulation, (2) multi-database searching, (3) automated literature snowballing, and (4) data management and export. The study is approved by the University IRB office. Full exploratory study researcher demographics can be found in Appendix \ref{appendix:formative-demograpghic}.

During the interviews, we explored participants' current SLR workflows, challenges, and their expectations for an ideal automated support tool. We analyzed the interview transcripts using Reflexive Thematic Analysis \cite{braun2019reflecting}. The analysis focused on identifying recurring challenges and latent needs that could directly inform the design of \sysname{}. We acknowledge our positionality: the first author brought a deep familiarity with SLR best practices, while the second author brought an `outsider' perspective with a strong background in qualitative methods, enabling us to challenge assumptions and ground our themes in participant data.

\subsection{Challenges Identified in Modern SLR Workflows}

\subsubsection{Challenge 1 (C1): High Friction in Iterative Search Exploration.}
Researchers universally described SLR searching not as a single, linear task but as a deeply iterative and exploratory process. However, current tools impose significant friction on this process. A primary source of this friction stems from the need to query multiple, disparate academic databases, each with its own proprietary and inconsistent search syntax. As illustrated in Figure \ref{fig:formative-database-usage}, the researchers we interviewed rely on a wide array of sources. The most frequently mentioned databases were IEEE Xplore, ACM Digital Library, and Google Scholar, reflecting their central role in the research process. A major source of frustration is that ``every one of these platforms uses different syntaxes for the structure of the string'' (R18), a sentiment echoed by many (R1, R15, R17). This forces researchers to manually translate and tweak queries for each database, a tedious and error-prone task that fragments the workflow.

This cognitive burden is compounded by the challenge of managing keyword variations. Researchers must manually account for synonyms, regional spellings (e.g., R7's example of ``visuali\textbf{z}ation'' with an `\textbf{z}' or a `\textbf{s}'), and evolving terminology. As R1 noted, this iterative refinement is essential: ``It was an iterative process to determine what works.'' R1 also expressed a clear need for a feature that would allow them to see a ``diff to say if we tried this combination of keywords, these are the papers that we get... if we change the keyword by a little bit, then it removes these 40 papers that we don't care about.''

The desire to manage keyword variations and see synonyms directly relates to reducing extraneous cognitive load during query formulation, a key principle in designing effective search interfaces \cite{Sweller1988CognitiveLD, Hearst_2009}. Furthermore, the expressed need for comparing iterative searches resonates with theoretical frameworks of exploratory search and ``berrypicking'', where a search query is a dynamic hypothesis that shifts as the user gains knowledge from intermediate paper results \cite{Bates1989TheDO, 10.1145/1121949.1121979}. Supporting such dynamic, comparative workflows is crucial for empowering deep and rigorous exploration.

\subsubsection{Challenge 2 (C2): Overwhelming Scale and Pace of Literature.}
\begin{figure*}[ht]
    \centering
    \includegraphics[width=0.95\linewidth]{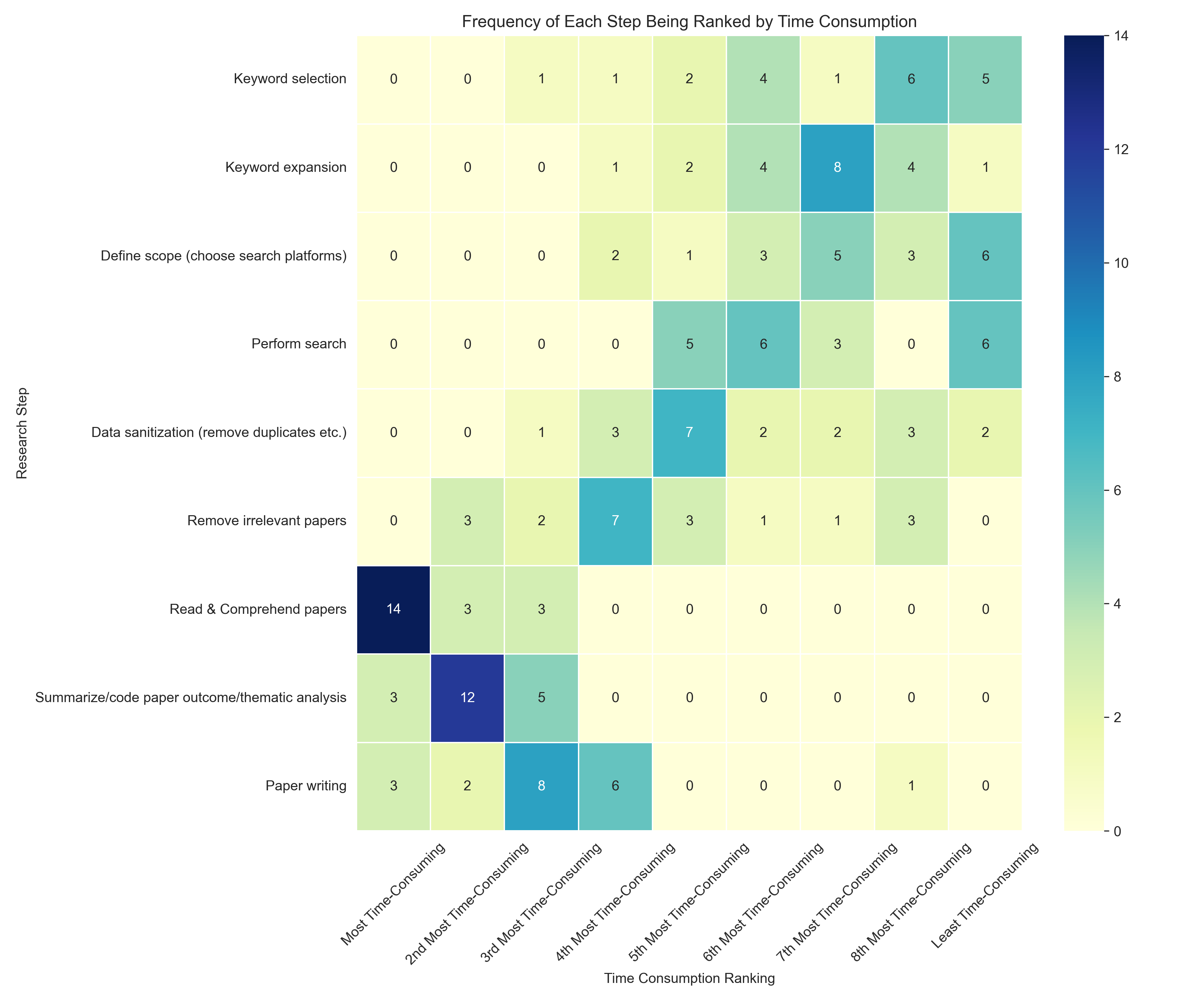}
    \caption{Heatmap of Research Steps Ranked by Perceived Time Consumption. This heatmap illustrates the frequency of rankings for nine different research steps based on their perceived time consumption, as assessed by interviewees.}
    \label{fig:interview-q5}
    \Description{This heatmap displays the frequency of rankings for nine research steps based on perceived time consumption, as evaluated by interviewees. The steps include keyword selection, keyword expansion, defining scope, performing searches, data sanitization, removing irrelevant papers, reading and comprehending papers, summarizing/coding paper outcomes, and paper writing. The color intensity corresponds to the number of interviewees who ranked each step according to how time-consuming it was, with darker colors indicating higher frequencies. "Read \& Comprehend papers" is perceived as the most time-consuming task, followed by "Summarize/code paper outcome" and "Paper writing"}
\end{figure*}

Researchers, particularly whom connected to fast-moving fields like AI, reported feeling overwhelmed by the sheer volume and relentless pace of new publications. This scale renders traditional review methods increasingly untenable. R9 captured this sentiment vividly: ``Generative AI and CSEd. There are hundreds of [papers]... my techniques that I would have used in the past for lit views just wouldn't work.'' The rapid obsolescence of knowledge further complicates matters; a literature review conducted in August was described as ``hilariously out of date already'' by the following April.

This explosion of publications places an immense labor burden on researchers across the entire review lifecycle. Our survey of time-consuming tasks (Figure \ref{fig:interview-q5}) reveals a critical distinction in this burden. While the core intellectual work of reading, comprehending, and synthesizing papers was, as expected, ranked as the most time-consuming steps, the data also highlights that a noticeable portion of researchers' time is consumed by a series of procedural bottlenecks that precede this high-level analysis.

Tasks such as keyword expansion, data sanitization, and particularly the initial filtering of irrelevant papers, were all identified as major time sinks. A significant portion of this procedural labor is spent on what R10 described as the necessity to ``go through a lot of irrelevant papers'' to ensure comprehensive coverage. R13 elaborated on this manual effort, recounting the need to ``handwork through thousands and thousands of records to check everything.'' This presents a clear opportunity for intelligent systems to assist in excluding irrelevant literature, thereby freeing researchers to focus on knowledge synthesis.

\subsubsection{Challenge 3  (C3): Balancing Automation with Scholarly Agency.}
Faced with information overload, researchers are increasingly turning to generative AI for assistance. However, this adoption is accompanied by significant concerns about reliability, accuracy, and the preservation of scholarly autonomy. While acknowledging the competitive need for AI assistance, as R9 put it, ``It's like GenAI versus me, and I'm never going to win that'', participants stressed that critical thinking cannot be outsourced. R4 insisted that ``comprehending papers is independent of a tool. You have to do it yourself.''

This tension gives rise to a clear expectation for how automation should be designed: not as an opaque black box, but as a transparent, controllable, and verifiable partner. The ideal role for AI is to assist with laborious tasks like initial filtering, but only if the researcher remains in control. R10 powerfully articulated this principle of verifiable assistance: ``I'd want it to be able to verify the things it assisted us... verifying is often quicker than doing it yourself. but I wouldn't want it to be like in the background, because then I maybe wouldn't trust the resource fully.''

This user expectation aligns directly with the core principles of human-centered AI   \cite{shneiderman2020humancenteredartificialintelligencereliable} and echoes recent calls for SLR tools that explicitly prioritize transparency and user control, thereby promoting appropriate trust in automated systems \cite{10.1145/3706598.3714047}. To facilitate appropriate reliance and trust \cite{lee2004trust}, systems must be transparent and give users ultimate control \cite{shneiderman1997direct}. Techniques like providing LLM reasoning for its suggestions \cite{10.1145/3613904.3641960} and using few-shot prompting from user-provided examples \cite{NEURIPS2020_1457c0d6} are practical design patterns that empower users, reinforce their agency, and align with guidelines for building trustworthy human-centered AI systems \cite{shneiderman2020humancenteredartificialintelligencereliable, Wiberg:2023aa, 10.1145/3643491.3660283}.\enlargethispage*{12pt}

\subsection{Design Goals}
\label{sec:design-goal}


\begin{table*}[h!]
\centering
\begin{tabular}{p{0.25\linewidth} p{0.25\linewidth} p{0.4\linewidth}}
\toprule
\textbf{Challenge (C)} & \textbf{Observed Researcher Need (N)} & \textbf{Design Goal (DG) \& System Feature (F)} \\
\midrule
\textbf{C1.} High Friction in Iterative Search Exploration & 
\textbf{N1.} Manage keyword variations and compare search iterations without manual overhead. & 
\textbf{DG1. Support Fluid, Iterative Exploration} \newline $\rightarrow$ \textbf{(F1)} Keyword Variation Management \newline $\rightarrow$ \textbf{(F2)} Iterative Search Comparison \\
\hline
\textbf{C2.} Overwhelming Scale and Pace of Literature & 
\textbf{N2.} Reduce the manual labor of filtering thousands of irrelevant papers. & 
\textbf{DG2. Mitigate Information Overload} \newline $\rightarrow$ \textbf{(F3)} AI-Assisted Filtering \\
\hline
\textbf{C3.} Balancing Automation with Scholarly Agency & 
\textbf{N3.} Leverage AI for efficiency while retaining control and the ability to verify automated suggestions. & 
\textbf{DG3. Uphold Agency via Verifiable AI} \newline $\rightarrow$ \textbf{(F3)} The design of the AI-Assisted Filtering feature, specifically: \newline \quad \textbf{(F3a)} User-guided calibration \newline \quad \textbf{(F3b)} AI provides reasoning for tag\\
\bottomrule
\end{tabular}
\caption{Mapping from user challenges and behaviours to design goals and system features. The feature labels (F1, F2, F3) correspond to the subsections in Section 4.}
\label{tab:challenges-to-goals}
\end{table*}


Drawing from the challenges identified in our exploratory study, we formulated three core design goals to guide the design and development of \sysname{}. These goals aim to directly address the challenges and needs of researchers, grounding our system in a human-centered approach to supporting systematic literature reviews. Table \ref{tab:challenges-to-goals} maps our observed challenges to the corresponding design goals and their associated system features.

\begin{enumerate}
    \item \textbf{Design Goal 1 (DG1): \emph{Support Fluid, Iterative Search Exploration}.} To address the challenge of high friction in current SLR workflows (C1), our primary goal is to create an integrated environment that streamlines the iterative and exploratory nature of searching. This involves automating the management of keyword variations (F1) and providing tools for direct comparison between search iterations (F2), thereby reducing cognitive load and empowering researchers to refine their strategies efficiently.
    \item \textbf{Design Goal 2 (DG2): \emph{Mitigate Information Overload with Controllable Automation}.} To tackle the overwhelming scale and pace of modern literature (C2), we aim to leverage automation to reduce the most laborious and time-consuming aspects of the SLR process through an AI-assisted filtering feature (F3). The goal is to shift the researcher's effort from tedious manual exclusion to higher-level knowledge synthesis and analysis.
    \item \textbf{Design Goal 3 (DG3): \emph{Uphold Researcher Agency through Transparent, Verifiable AI}.} To resolve the tension between automation and scholarly agency (C3), our goal is to ensure our AI-powered assistance (F3) is transparent, controllable, and verifiable. Instead of replacing human judgment, the system must augment it by providing suggestions with clear rationales (F3a) and ensuring the user agency (F3b).
\end{enumerate}



\section{\sysname{}: A Design Probe to Investigate the SLR Workflow}
Guided by the findings from our exploratory design study in Section \ref{sec:design-goal}, we developed \sysname{} as a high-fidelity design probe. Our design philosophy was grounded in established principles of mixed-initiative system design \cite{10.1145/302979.303030}, which envisions a flexible interaction where control can shift dynamically between the user and the system based on context and capability; and human-AI collaboration \cite{10.1145/302979.303030, 10.1145/3290605.3300233}, which advocate for keeping the user in control and promoting appropriate reliance with AI. This section details how we operationalized our three design goals into a cohesive, web-based tool. Each core feature was designed as a targeted approach to address a specific, observed challenge, allowing us to evaluate its impact on the scholarly review workflow.

\subsection{{\raisebox{-0.5ex}{%
 \protect\includegraphics[height=1em]{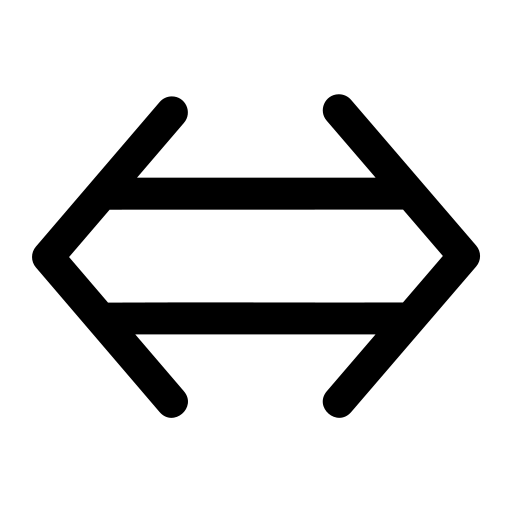}} %
  Keyword Variation Management (F1)}}

Keyword Variation Management (F1), targets the high cognitive load of managing syntax. Participants in our exploratory study noted the tedious need to manually account for synonyms and regional spellings (e.g., R7). To investigate whether automating this syntactic labor could free up cognitive resources for conceptual work, our probe integrates keyword assistance that helps researchers identify more relevant keywords.

As a researcher types a search query, \sysname{} automatically suggests relevant synonyms and common spelling variants (e.g., British vs. American English) directly within the interface. These suggestions are generated using a combination of lexical databases (breame \cite{breame2023}) and synonyms API (Thesaurus \cite{thesaurus2024}). Researchers can easily add these variations to their search string, which the system automatically formats with the correct Boolean operators. By automating this manual task, \sysname{} aims to reduce extraneous cognitive load, allowing researchers to focus on the semantic formulation of their search strategy rather than its syntactic implementation.

\subsection{{\raisebox{-0.5ex}{%
  \protect \includegraphics[height=1em]{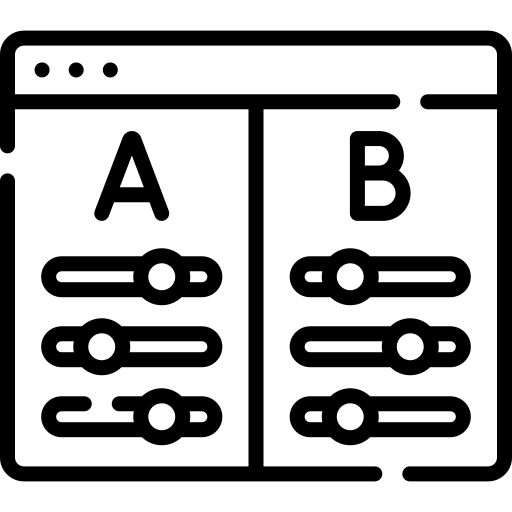}} %
Iterative Search Comparison (F2)}}
\begin{figure}
    \centering
    \includegraphics[width=0.95\linewidth]{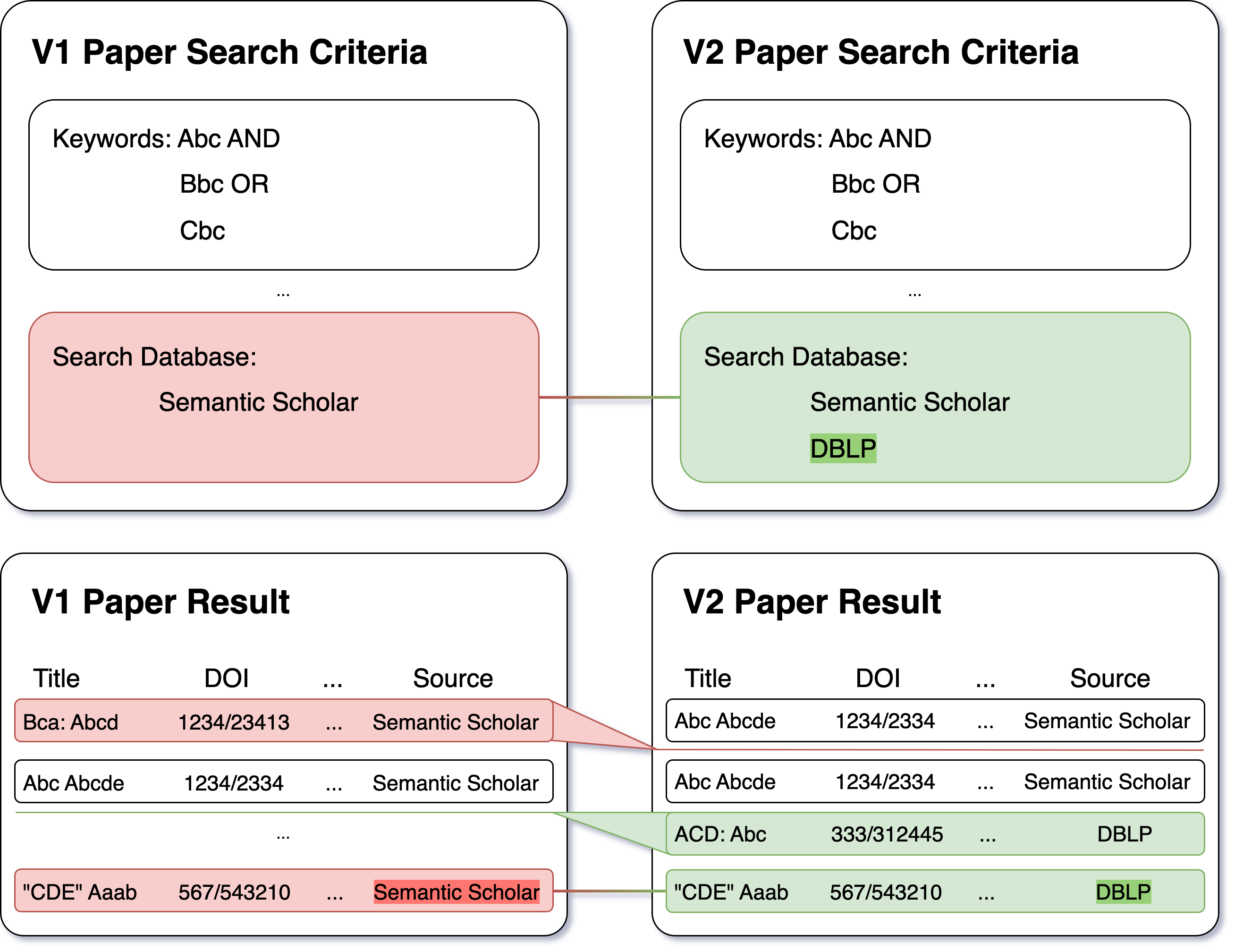}
    \caption{Iterative Search Comparison Feature (F2). Users are able to select any two searches they performed and directly compare the differences between them. Differences are shown in both search criteria (i.e., the difference in keywords, scholar databases) and resulting papers.}
    \label{fig:diff-function}
    \Description{This figure shows \sysname's search comparison feature, enabling users to compare two separate searches (V1 and V2). The top section displays the search criteria for both versions, where V1 uses only the Semantic Scholar database, while V2 adds DBLP to the search. The bottom section shows the paper results for each search. The comparison highlights differences in the resulting papers, including variations in sources (Semantic Scholar vs. DBLP) and paper titles. The feature allows users to visually track and understand differences between search criteria and resulting papers across databases.}
\end{figure}

We implemented Iterative Search Comparison (F2) to support transparent refinement (DG1), a direct response to participant requests for a ``diff'' tool (R1). This feature operationalizes query refinement as a comparative experiment rather than a linear history. It allows researchers to audit the impact of their adjustments by calculating the exact delta between search versions, isolating which papers were added or lost, to shift the workflow from trial-and-error to strategic, evidence-driven exploration.

As shown in Figure \ref{fig:diff-function}, \sysname{} enables researchers to select any two search attempts from their history and view a side-by-side comparison. The interface clearly highlights differences in both the search criteria (e.g., added or removed keywords, different databases selected) and, crucially, the resulting set of retrieved papers. It explicitly lists which papers were added, which were removed, and which remained between the two versions. This feature transforms the abstract process of query refinement into a concrete, auditable activity. It aims to empower researchers to make more informed decisions about their search strategy, ensuring that each modification is a deliberate step toward a more precise and comprehensive set of results.

\subsection{{\raisebox{-0.5ex}{%
  \protect\includegraphics[height=1em]{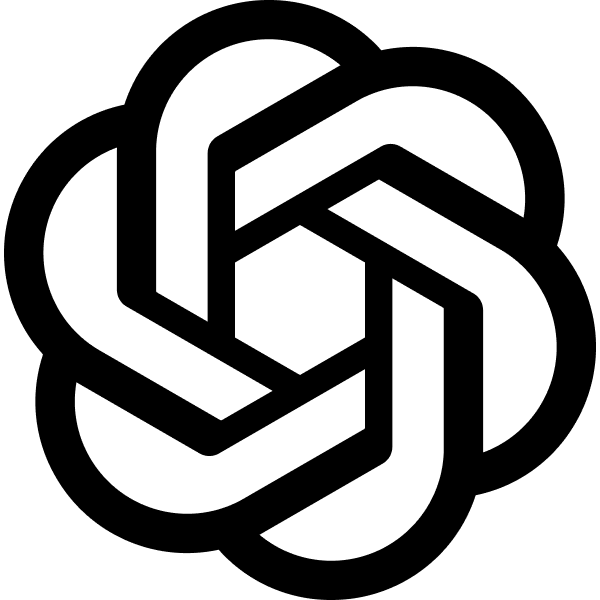}} %
AI-Assisted Irrelevance Filtering (F3)}}
Our final major feature was designed to explore the tension between automation and agency (DG2 \& DG3). The exploratory study revealed that while filtering was a major bottleneck, researchers were deeply concerned about the opacity of AI systems. We therefore implemented a verifiable, human-in-the-loop filtering system to investigate whether such a partnership could reduce manual labor while preserving scholarly control. This feature extends in a two-step process designed to keep the researcher in command:

\subsubsection{User-Guided Calibration (F3a):} The process begins with the researcher defining their own analytical framework directly within the system. First, they input a specific inclusion/exclusion criterion they wish to evaluate (e.g., \code{Study Methodology}, \\\code{Population Focus}). Next, they define the custom set of possible labels for that criterion. For instance, for the \code{Study Methodology} criterion, a researcher might define the labels as \nestedcode{Quantitative}, \nestedcode{Qualitative}, \nestedcode{Mixed-Methods}, or \nestedcode{Review Paper}.

The researcher initializes the calibration process by annotating a small seed set of papers (typically $N=3$). For each instance, they assign a label from their defined schema and articulate a concise rationale derived from the abstract and metadata. This design serves two distinct functions. Technically, it generates structured few-shot examples that condition the underlying LLM (OpenAI’s \texttt{o3-mini} \cite{openai2025o3mini}) to the specific decision logic. Interactionally, it allows the researcher to explicitly encode their subjective analytical boundaries, ensuring the subsequent automation operates within the user's specific scholarly context rather than applying generic relevance criteria. The full system prompt is detailed in Appendix \ref{sec:appendix-prompt-for-llm-filtering}.

\subsubsection{Suggestion with Rationale (F3b):} Once received with the researcher's guided examples, the system then automatically screens the remaining unclassified papers against the user-defined criterion. For each paper, it suggests the most appropriate label from the researcher's custom-defined set.

Additionally, this suggestion is always accompanied by a concise, evidence-based rationale explaining why it selected that specific tag. For example, continuing with the \code{Study Methodology} criterion, the system might process a new paper and suggest the label \nestedcode{Quantitative}, providing the rationale (i.e., \emph{``The abstract describes the use of `statistical analysis' and a `controlled experiment with 250 participants,' which aligns with a \nestedcode{quantitative} methodology''}). This transparency allows the researcher to quickly assess the validity of the AI's reasoning before accepting or overriding the suggestion.

\subsection{System Implementation and Architecture}
To ensure that our user evaluation would generate valid findings based on a realistic and reliable experience, we built \sysname{} as a robust, full-functioning web application. The following architectural and implementation choices were essential supporting infrastructure of our core design goal.

\subsubsection{{\raisebox{-0.5ex}{
   \protect\includegraphics[height=1em]{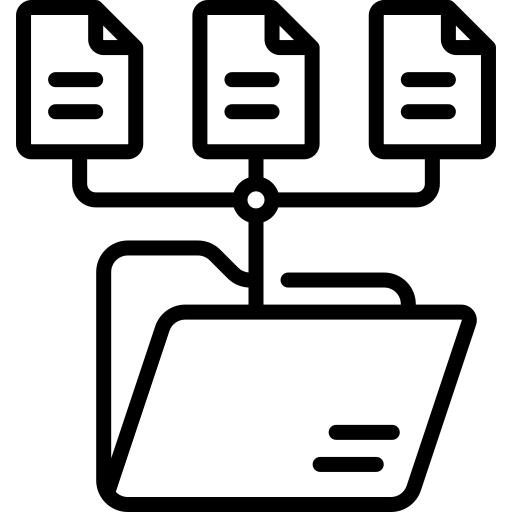}} %
Multi-Database Search Pipeline}}
To mitigate publisher bias and maximize search comprehensiveness \cite{10.1007/s11192-018-2958-5}, \sysname{} integrates with multiple academic databases. Users can select and query any combination of sources, including \textit{Semantic Scholar}, \textit{DBLP}, \textit{Web of Science}, \textit{IEEE Xplore}, and \textit{Scopus}. Our selection was guided by each source's coverage, API availability, and researcher practices (as seen in the exploratory study and Figure \ref{fig:formative-database-usage}). Semantic Scholar is utilized for keyword searching, metadata extraction, and snowballing, while IEEE, DBLP, and Web of Science primarily serve as search sources. For robust metadata population, we also integrate \textit{Crossref}, the world’s largest registry of DOIs \cite{10.1162/qss_a_00112}. We deliberately excluded Google Scholar due to its lack of a first-party API, inconsistent inclusion of DOIs in results, and documented limitations in precision, transparency, and reproducibility for systematic reviews \cite{https://doi.org/10.1002/jrsm.1378}.

\sysname{} supports two primary keyword structures to accommodate different research practices. First, users can provide an \textit{advanced search string}, enabling complex Boolean logic (e.g., \texttt{"A" AND "B" OR "C" NOT "D"}) and compatibility with established frameworks like PICO \cite{Schardt:2007aa} or SPICE \cite{Booth:2006aa}. Second, \sysname{} supports \textit{multi-layer keyword lists}, where users input terms into separate groups. The system then automatically generates a cross-product of these terms linked with \texttt{AND}, equivalent to a query like \texttt{("A" OR "B") AND ("C" OR "D")}, a structure common in SLR protocols \cite{omer2021introductory, GAROUSI2016106}.

Beyond keyword searching, \sysname{} implements forward and backward reference searching (also known as snowballing), a critical technique for comprehensive literature discovery \cite{10.1145/2372251.2372257, PMID:25145803}. Our system adopts a free-form approach \cite{WOHLIN2022106908}, allowing researchers to dynamically initiate snowballing not just from an initial seed set, but from any paper within the system at any stage of the review process.

\subsubsection{{\raisebox{-0.5ex}{%
   \protect\includegraphics[height=1em]{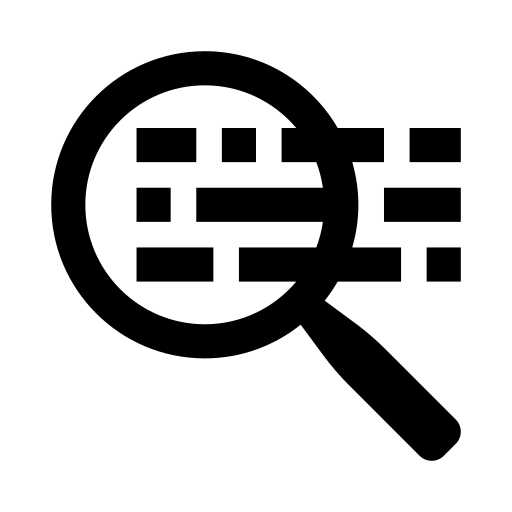}} %
   Metadata Curation and Workflow Support}}
Once papers are retrieved, \sysname{} performs several automated curation tasks. To support filtering and classification \cite{10.1186/2046-4053-4-5, Timothy}, the system populates a rich set of metadata for each entry, including title, authors, affiliations, venue, DOI, abstract, keywords, publisher, and publication date. As papers are collected from multiple sources, the system automatically removes duplicates based on their unique Digital Object Identifier (DOI) and title, ensuring a clean dataset.

To support search validation and workflow interoperability, we integrated two key features. First, researchers can input a list of known, expected papers, and \sysname{} will automatically calculate and display the retrieval percentage, offering a rapid method for assessing search quality (a practice used by R17). Second, recognizing that researchers rely on an ecosystem of external tools like Zotero, EndNote, and Mendeley, \sysname{} is designed for seamless data exchange. To facilitate this, users can initiate analysis by importing an existing library, or export their curated collections in standard formats, including \texttt{.csv} for spreadsheet processing, \texttt{.bib} for scientific writing \LaTeX, and \texttt{.ris} for reference managers. This ensures that \sysname{} acts as a fluid bridge in the sensemaking process rather than a siloed application.



\section{Comparative User Study}
To evaluate the impact of these features from \sysname{}, we conducted a formal comparative user study. The goal of the study was to investigate how an integrated, human-in-the-loop environment impacts researchers' strategies, cognitive load, and confidence compared to the baseline condition.

\subsection{Participants}
We recruited eight participants with prior knowledge and experience of systematic literature review (P1–P8; six men, one woman, one undisclosed) through internal mailing lists and social media. The cohort consisted primarily of Ph.D. students (6/8), one M.Sc. student, and one assistant professor, all of whom had a background in computing. Participants possessed significant research experience (M=4.88 years, SD=3.80) and reported high familiarity with the SLR process (M=4.37/5) and were generally familiar with baseline tools (Google Scholar: M=4.25/5; Google Sheets: M=3.88/5). Each participant was compensated \$20 for their time, and the study protocol was approved by our institution's research ethics board. Full comparative user study participant demographic and self-report experience with SLR can be found in Appendix \ref{appendix:user-eval-demograpghic}.

\subsection{Tasks and Baselines}
The study was structured around three fundamental SLR tasks. For each task, participants used both \sysname{} and a corresponding baseline tool.

First, for \textbf{paper searching}, participants were asked to locate relevant literature within a given research context. To simulate the iterative nature of real-world SLRs, the search task was phased, simulating the iterative nature of real-world research, that participants refine their search keywords as they progressed. The baseline for this task was Google Scholar, a widely-used academic search engine~\cite{falagas2008comparison}.

Second, for \textbf{snowballing}, participants performed both forward (citations) and backward (references) searches on a seed paper to identify connected works~\cite{Greenhalgh1064, WebsterWatson2002, WOHLIN2022106908}. The baseline combined Google Scholar for its citation-tracking features and Google Sheets for managing the collected papers, a common practice in collaborative SLRs~\cite{carrera2022conduct}.

Third, for \textbf{paper filtering}, participants were tasked with applying one inclusion-exclusion criterion to a list of 15 papers based on their titles and abstracts, following PRISMA guidelines~\cite{Pagen71}. The baseline for this task involved using Google Sheets for organization alongside the publisher's website for accessing abstracts.

To mitigate carry-over and learning effects in our within-subject design~\cite{CHARNESS20121}, we created two distinct SLR scenarios: (1) \textit{AI in Education}, inspired by the work of \citet{10.1145/3623762.3633499}, and (2) \textit{Mental Health Chatbots}, based on research by \citet{abd2019overview, abd2021perceptions}. The order of the SLR setting was fixed (i.e., we always started with \textit{AI in Education}, then transitioned to \textit{Mental Health Chatbots}), while the tool condition for each task was randomized (i.e., start with \sysname{} condition vs. start with baseline condition).

\subsection{Rationale for Baseline Setup}
To ensure ecological validity, we selected a baseline that reflects the predominant workflow of researchers identified in our exploratory design study and other prior literature. While specialized tools like Rayyan \cite{rayyan} or Paper Fetcher \cite{paperfetcher} exist, our initial interviews and prior literature \cite{afifi2023data} confirm that the vast majority of researchers still rely on a combination of general-purpose tools. Consequently, we utilized Google Scholar \cite{falagas2008comparison, sadeghi2024systematic} for retrieval and Google Sheets for manual organization. As there is no single, standard automated toolchain, any attempt to impose a specialized tool as a baseline would create an artificial scenario. This would confound the study with the cognitive overhead of learning a new baseline tool, hindering our ability to yield meaningful insights into how \sysname{} improves upon real-world analytical practice. We utilized Google Scholar as a common interface for paper retrieval, not as a standard for reproducibility, but as a reflection of common practice.

\subsection{Procedure}
Each one-hour study session was conducted remotely via video conference. After providing informed consent and completing a demographic survey, participants were randomly assigned to one of two groups. The first group completed the searching task, while the second group completed the snowballing and filtering tasks sequentially. To reduce cognitive fatigue, a five-minute break was provided between the snowballing and filtering tasks. For each assigned task, participants completed it once using \sysname{}, under a pre-set scenario (i.e, \textit{AI in Education}) and once using the baseline tool, with a different scenario (i.e., \textit{Mental Health Chatbots}). The order of tool exposure was counterbalanced. Following the completion of each task under both conditions, participants provided ratings on their experience. The study concluded with a survey and a semi-structured interview to collect overall feedback and detailed qualitative insights. The full study procedure is illustrated in Appendix \ref{appendix:user-study-procedure}.

\subsection{Measures and Analysis}

To quantitatively assess perceived workload, we administered the NASA-TLX after each task completion \cite{hart2006nasa}, alongside ratings of \textit{assurance} (subjective confidence and informedness) within each phase of the search task. Additionally, to evaluate the holistic user experience, we administered a 12-item post-study questionnaire (using a 7-point Likert scale) that assessed five key dimensions: efficiency, interface clarity, workflow guidance, confidence, and overall satisfaction.

To gather qualitative data on user experience, pain points, and perceived benefits, we conducted a semi-structured interview at the end of the session. The qualitative data from these interviews were thematically analyzed to identify recurring patterns and insights regarding the system's utility and usability compared to existing workflows.

Additionally, we recorded objective performance metrics for the filtering task to assess the burden of human verification. We compared \textit{Manual Time} against \textit{AI Generation Time} plus \textit{Human Verification Time}, and calculated accuracy against a ground-truth inclusion set. To establish this ground truth, the first and second authors iteratively discussed the inclusion criteria and independently coded the dataset, achieving perfect inter-rater agreement. We focused on objective metrics on filtering, as search and snowballing comprehensiveness are functionally dependent on external APIs rather than the interface design.

For qualitative interview data, we applied Reflexive Thematic Analysis \cite{braun2019reflecting} and followed a similar procedure to that outlined in Section \ref{sec:formative-design} for the exploratory design study.

\section{Result}
Full quantitative statistics can be found in Appendix \ref{appendix:user-study-stats}.

\begin{figure*}[h!]
    \centering
    \includegraphics[width=0.96\linewidth]{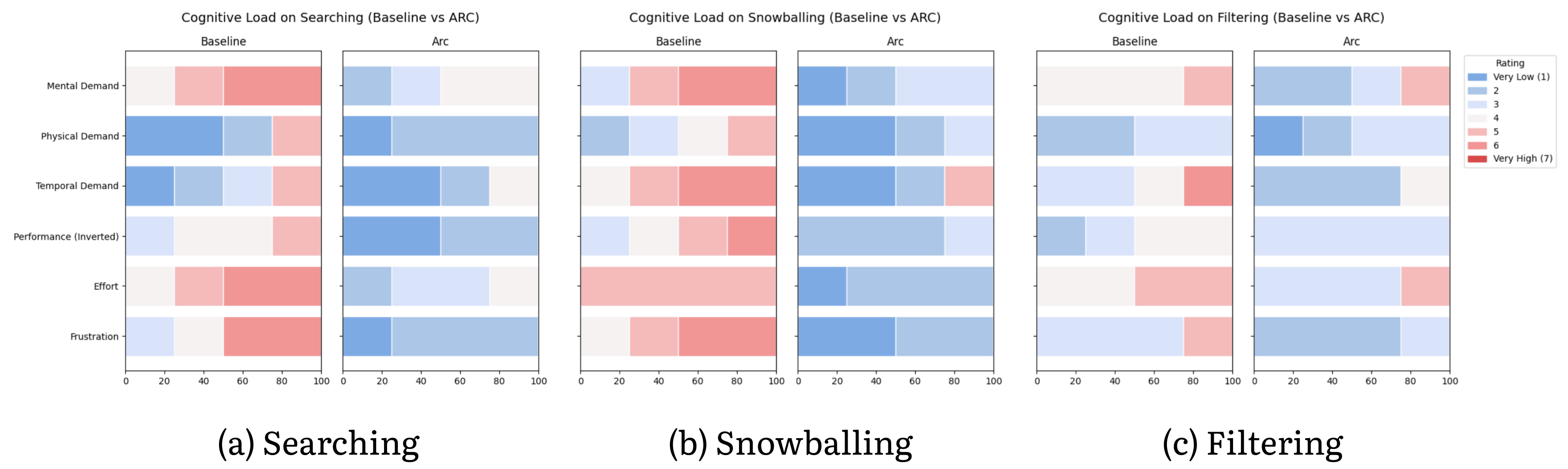}
    \caption{Cognitive Load on User Study Tasks.}
    \Description{
    Three diverging stacked bar charts compare the cognitive load between the Baseline condition and the Arc condition across three tasks: Searching, Snowballing, and Filtering. The charts measure dimensions including Mental Demand, Physical Demand, Temporal Demand, Performance, Effort, and Frustration. For the Searching task, the loads are somewhat comparable but lean lower for Arc. However, for the Snowballing and Filtering tasks, there is a distinct difference; the Arc bars show significantly more blue, indicating lower perceived workload and frustration, compared to the red-heavy bars of the Baseline condition.
    }
    \label{fig:cog-load}
\end{figure*}

\begin{figure}[h!]
    \centering
    \includegraphics[width=0.96\linewidth]{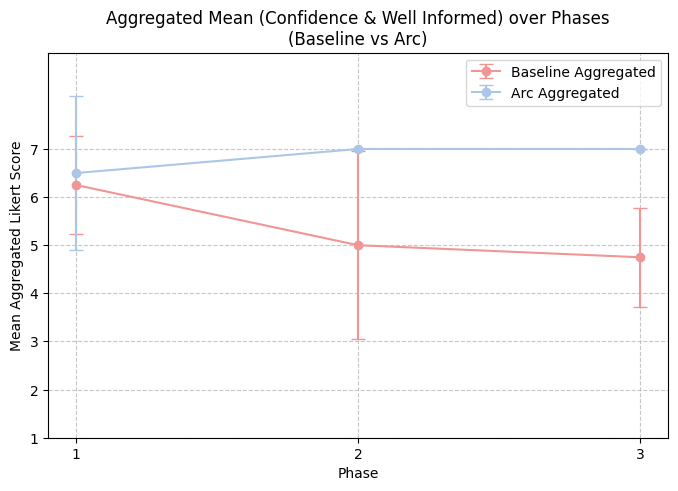}
    \caption{Aggregated Assurance Across Iterative Searching Phases. We measure assurance via separate questions on subjective confidence and well-informedness of participants at the end of each searching phase.}
    \Description{A line graph plotting the Mean Aggregated Assurance score across three phases of the iterative searching task. The Baseline condition, shown in red, starts with a score around 6.25 but steadily declines to approximately 4.75 by Phase 3. The Arc condition, shown in blue, starts slightly higher and increases to reach the maximum score of 7 by Phase 2, maintaining that high level through Phase 3. Error bars indicate the standard error for each data point.}
    \label{fig:searching-phases}
\end{figure}

\subsection{Insight 1: Transparent Comparison Enables Systematic Query Refinement}
\label{sec:results-searching}

Addressing RQ1, we observed that \sysname{}'s interface encouraged participants to move from memory-based iterations to systematic experimentation. This was reflected in perceived cognitive load, where the mean NASA-TLX score was lower when performing search using \sysname{} compared to the baseline (Baseline: $M=4.04$, $SD=0.71$ vs \sysname{} $M=2.21$, $SD=0.44$. See Figure \ref{fig:cog-load}a). Participants attributed this improvement to the integrated workflow, which enabled a more strategic approach to refinement.

Participants frequently highlighted the iterative search comparison (``\texttt{diff}'' view) as a mechanism for understanding their search process (P1, P2, P4). By providing a visual log of how query modifications affected the result set, the feature helped address the ``noise'' of unmanageable results (P1). P3 noted that this visual feedback helped validate hypotheses about keyword changes.

The ability to compare results supported a change in search strategy. P2 described this as moving from an inefficient ``depth-first search'' to a systematic ``breadth-first'' exploration. The persistent search history acted as a ``scratch pad'' (P2), removing the fear of losing promising results. This contrasted with the baseline, where participants reported frustration with opaque search logic (P1) and unpredictable sorting algorithm (P2).

Participants noted that the integrated history offloaded the labor of external documentation. P3 explained that \sysname{} eliminated the need to manually track queries, allowing them to focus on formulation:
\begin{quote}
\textit{``I don't feel like I need to separate my brain, leaving half of them trying to remember [the queries]. I can use all of them [with \sysname{}] just trying to explore what formula work and does not... it's a workflow in my brain, I think, that is improved.''} (P3)
\end{quote}

Participant assurance levels diverged over the course of the tasks (\autoref{fig:searching-phases}). While initially comparable (Baseline: $M=6.25$; \sysname{}: $M=6.50$), assurance in the baseline condition decreased as complexity increased. By the final phase, assurance level in the baseline dropped ($M=4.75$) while remaining high for \sysname{} ($M=7.00$), suggesting that the system helped sustain confidence and keep the user well-informed during complex search tasks.

\subsection{Insight 2: Automating Logistics Reduces Workload and Increases Efficiency}

Addressing RQ2, the snowballing task showed distinct differences in efficiency. The manual baseline required an average of 10.68 minutes ($SD=2.0$) to complete, while participants completed the same task using \sysname{} in less than one minute. This efficiency gain was mirrored in perceived workload, the overall NASA-TLX score decreased from a baseline mean of 4.75 ($SD=0.69$) to 1.96 ($SD=0.74$) with \sysname{} (see Figure \ref{fig:cog-load}b). Participants described the baseline process as a bottleneck. The inefficiency of manually transferring data prompted one participant to ask, ``Is there any fast method?'' (P7). Others described the manual process as slow (P5) and frustrating (P8).

Conversely, \sysname{}’s integrated and automated approach was seen as both novel and highly useful. Participants could seamlessly execute both forward (cited by) and backward (references) searches, with results presented in a unified, color-coded interface. The value of this automation extended beyond speed, participants felt it made the process more reliable and trustworthy by minimizing the potential for human error. P5 articulated this unique benefit:
\begin{quote}
    \textit{``I think [it makes] me much safer... because if I do its operations manually, there could be mistakes right''} (P5)
\end{quote}
The system's ability to produce a clean, unified, and de-duplicated dataset was seen as a key advantage that supports subsequent collaborative review stages by removing the need for a separate, error-prone data-cleaning step. This was valued even by experienced researchers who already use custom scripts to automate parts of their workflow, with P6 noting, ``\sysname{} is basically doing that for me, which is good'' (P6).

\subsection{Insight 3: Verifiable AI Elevates the Researcher's Role from Manual Screener to Strategist}
Addressing RQ3, our findings suggest that the human-AI partnership offloads manual screening, allowing researchers to focus on verification. Participants reported reduced cognitive load when using \sysname{} for filtering (Baseline $M=3.67$, $SD=0.36$ vs. \sysname{} $M=2.75$, $SD=0.21$. See Figure \ref{fig:cog-load}c).

Our observational data support this role shift with objective efficiency gains. For the 15-paper screening task, manual filtering took an average of 10:58 minutes, whereas the AI-assisted workflow (combining an average of 2 minutes of AI filtering and an average of 7 minutes of human verification) took approximately 9 minutes, an 18\% reduction in total effort. Crucially, the verification time alone (7 mins) was notably shorter than the manual baseline. Furthermore, the human-AI partnership appeared to reduce human error. In the manual baseline, participants correctly classified 75\% of papers. In contrast, the AI correctly categorized 95\% of the papers, allowing these same participants to focus their efforts on verifying the AI's rationales and self-correcting the remaining errors.

Participants attributed efficiency gains to the centralized interface. The baseline method required context-switching between tools, whereas \sysname{}'s unified view presented the title, abstract, and controls together. P7 noted:
\begin{quote}
    \textit{``...having everything in one place... instead of going to each paper each time, each abstract. Opening window closing window, a new tab so that's definitely, one big advantage.''} (P7)
\end{quote}
This consolidation allowed researchers to move from low-level manual data handling to a higher-level task of supervision. The tool's impact was framed as a direct time-saver, with one participant estimating that for a large pool of papers, \sysname{} could \textit{``reduce a significant amount of time''} (P5).

Participants found \sysname{}’s human-in-the-loop approach to automation, where they calibrate the AI with a few examples, to be both effective and empowering. P5 appreciated this, noting, ``the way that I can train a model to do... inclusion for me. So I think that's great'' (P5). For experienced researchers, this represented a conceptual leap from their existing methods. P6 contrasted \sysname{}’s capabilities with his prior, less sophisticated workflow: ``The way that I used to do filtering is that once I scraped everything... I did stuff like fuzzy search. And different keywords... Because I was not using any LLM. Any reasoning or inference to filter'' (P6).

Despite satisfaction with AI accuracy, participants emphasized the need of human verification. P8 noted a case where the AI label was correct, but the rationale did not align with expert heuristics. Expressing scholarly caution, P6 stated, ``I can never be certain that it will be correct all the time.'' Consequently, every participant engaged in a thorough verification of the AI's suggestions.

Finally, the study revealed tensions between \sysname{}'s structured workflow and the more dynamic, iterative nature of real-world systematic reviews. P8, a university professor, noted that filtering criteria are often not static but are refined through team negotiation as new types of papers are encountered. He observed that the current system did not support this emergent process, stating, ``usually at least during the pilot, there are like negotiations of what goes in and what goes out, which change the schema'' (P8). Similarly, other participants noted that their real-world SLRs require multiple screeners to ensure inter-rater reliability, a collaborative workflow the current individual-focused interface does not yet address (P7).

\subsection{Overall User Experience: A More Strategic and Confident Workflow}

\begin{figure*}[h!]
    \centering
    \includegraphics[width=0.96\linewidth]{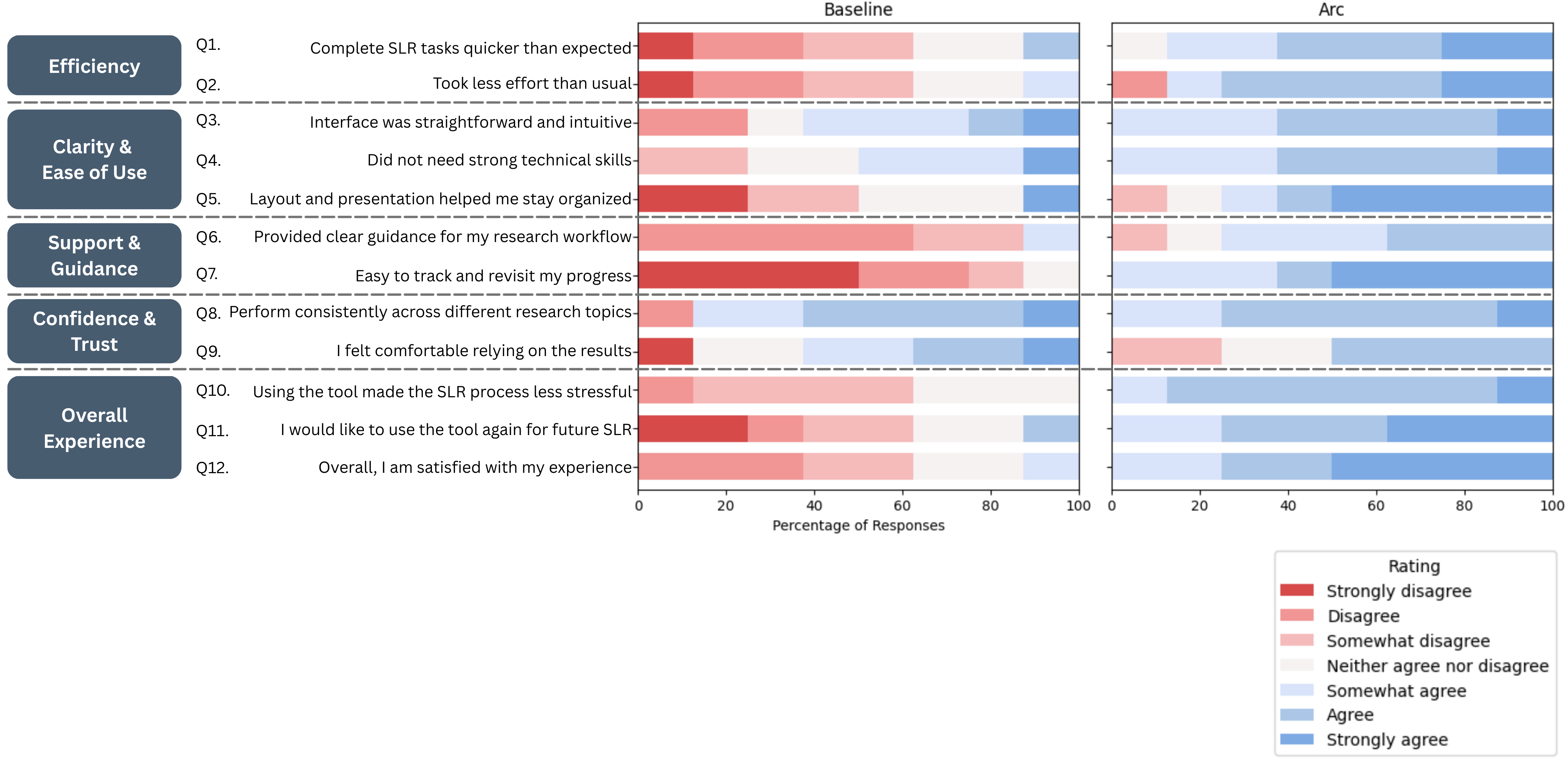}
    \caption{End of Study Survey.}
    \Description{A stacked bar chart visualizing responses to twelve post-study survey questions comparing the Baseline and Arc systems. Responses range from "Strongly Disagree" in red to "Strongly Agree" in blue. The Arc bars are overwhelmingly blue, indicating high user satisfaction across all categories including efficiency, interface clarity, and confidence. The Baseline bars show mixed results with significant red and pink portions, particularly for questions regarding the ease of tracking progress, stress levels, and overall satisfaction.}
    \label{fig:end-of-study-survey}
\end{figure*}

Taken together, these insights demonstrate how an integrated, human-in-the-loop environment fundamentally reframes the SLR workflow. The impact was indicated in participants' overall assessments (Figure \ref{fig:end-of-study-survey}, Q1-12) where the integrated workflow was rated more satisfactory ($M=3.12$ vs. $M=6.25$; Q12), less stressful ($M=3.25$ vs. $M=6.00$; Q10), and were more likely to adopt by participants for future tasks ($M=3.00$ vs. $M=6.12$; Q11). This preference was consistent across all dimensions. Qualitatively, participants attributed this positive experience to \sysname{}'s potential to integrate into and improve upon their established, real-world research workflows.

Beyond its specific features, participants also evaluated \sysname{}'s potential fit for real-world reviews. A key theme was the need for the tool to support the inherently collaborative and iterative nature of the research process. P7 and P8, both identified the current single-user design as a limitation, noting that real-world screening requires multiple researchers for inter-rater reliability and that filtering criteria often evolve through team negotiation. While others suggested small but impactful enhancements, such as keyboard shortcuts (P8) and bolding keywords in the filter rationale (P4), to improve usability and reduce friction. These findings highlight that while the core mechanics of \sysname{} were highly valued, its successful adoption depends on addressing these broader workflow and user experience needs.



\section{Discussion}

Systematic literature reviews are months-long projects, but they are built out of repeated, small operations: refining queries, tracking what changed, consolidating results across sources, screening at scale, and documenting decisions to meet reporting standards like PRISMA \cite{higgins2008cochrane, Pagen71, rethlefsen2021prisma}. Our comparative study evaluates a small slice of this work in a controlled setting ($N=8$), therefore, we use it to investigate whether an integrated, human-in-the-loop environment changes how individual researchers perform the high-frequency operations that dominate large reviews: whether they can iterate more deliberately, carry context forward with less mental overhead, and keep decisions auditable as complexity grows. With that lens, we discuss implications for tool design in three areas: shifting work from logistical coordination to strategic exploration; lowering barriers across fragmented databases; and supporting the systematic review lifecycle, from initial construction to long-term maintenance, followed by a call for more open academic infrastructure.

\subsection{From Toil to Thought: Offloading Logistical Friction to Enable Strategic Exploration}
Expert knowledge work is increasingly loaded by the cognitive burden of fragmented tools. Our study revealed that scholars spend significant intellectual energy on logistical tasks (translating syntax, de-duplicating records). This represents a misallocation of expertise, imposing extraneous cognitive load that limits capacity for the deep synthesis that is the unique provenance of the researcher \cite{Sweller1988CognitiveLD}.

\sysname{} demonstrates the potential for integrated environments to decouple strategic intent from mechanical execution. In current workflows, a researcher's mental model is often tightly coupled to the rigid syntax of specific databases; changing a theoretical direction requires a tedious manual refactoring of query strings. \sysname{} breaks this coupling. By externalizing the search process into a visual, manipulatable history, the system acts as a form of distributed cognition \cite{hollan2000distributed}, where the state of the inquiry is offloaded to the interface. This allows the researcher to manipulate the logic of the inquiry (e.g., ``what happens if I broaden the date range?'') while the system handles the implementation details. In our study, this shift appeared to foster a more systematic ``breadth-first'' exploration, a behavior often punished by the friction of traditional tools, suggesting that reducing the cost structure of these operations can potentially alter the strategies users employ \cite{russell1993cost}.

\subsection{Lowering Barriers to Support Interdisciplinary Knowledge Synthesis}
The current ecosystem of academic databases often functions as a collection of walled gardens. This fragmentation is well-documented, with studies revealing significant differences in coverage and retrieval quality across platforms, forcing researchers to navigate heterogeneous platforms and databases \cite{10.1007/s11192-018-2958-5, https://doi.org/10.1002/jrsm.1378}. This environment, where specialized search syntaxes and, more subtly, domain-specific terminologies create a high degree of epistemic friction, is a known bottleneck in the SLR process \cite{o2014techniques}. The burden of manually translating queries and concepts across platforms discourages exploration and creates uncertainty, inhibiting interdisciplinary discovery. Such barriers are a critical impediment to progress, as many of modern science's most complex challenges require the synthesis of knowledge from diverse fields \cite{stokols2008science}.

The design of \sysname{} embodies a generalizable pattern for bridging these knowledge silos. Its unified, multi-database search acts as an abstraction layer, hiding the idiosyncratic complexity of individual data sources behind a single, conceptually coherent interface. In our study, this abstraction allowed participants to shift from memory-based guessing to evidence-based validation of their queries. Crucially, features like keyword variation suggestions and transparent search history provide epistemic scaffolding for this exploration. By visualizing how specific keyword variations alter the result set (via the iterative \texttt{diff}), the system creates a tight feedback loop. While our participants were operating within their general domain, this mechanism, by making the semantic impact of syntactic changes visible, offers a potential pathway for interdisciplinary synthesis. It suggests that such interfaces can support a ``trial-and-error'' discovery process, allowing researchers to safely experiment with and validate the ``native'' keywords of an unfamiliar domain without needing to master its complex query syntax first.

In this architectural approach, the system acts as a ``universal translator'', allowing the researcher to use a single language of inquiry while the system handles the unique syntax of each database. This model of creating a unified layer over a fragmented data paradigm has applications far beyond academic literature review. The same principles could be applied to create integrated tools for legal research across disparate case law systems, for market analysis across proprietary data providers, or for intelligence gathering across heterogeneous information feeds. By providing a framework for unifying fragmented knowledge, a challenge also addressed in the context of collaborative sensemaking \cite{10.1145/1294211.1294215}, \sysname{} contributes a design pattern for the next generation of tools for thought.

\subsection{Sustainable Infrastructure for the Knowledge Review Lifecycle
} 
While our user study successfully demonstrated the immediate efficiency gains of using \sysname{} during initial search and screening, the true architectural significance of the system lies in its potential to support the long-term, multi-stage lifecycle of rigorous SLRs. Real-world high-quality reviews are not linear sprints, but prolonged, iterative endeavors often spanning months \cite{KITCHENHAM20132049s, WebsterWatson2002, higgins2008cochrane}. A primary friction point in this elongated process is the loss of strategic context. As Fok et al. highlight, the ``implicit strategies'' and subjective decisions driving a review are rarely documented, making it difficult for researchers to ``restore institutional knowledge'' when returning to update the work \cite{10.1145/3706598.3714047}. \sysname{} tackles this challenge by automatically logging the provenance of every search iteration and keyword modification, thereby transforming the ephemeral process of ``searching'' into a persistent, auditable asset. This ensures that as an SLR scales from hundreds to thousands of papers, the researcher does not lose their strategic foundation, effectively separating the intellectual overhead of the review from the cognitive load of remembering logistical details \cite{pirolli1999information, Sweller1988CognitiveLD}.

This state preservation is the critical facilitator for the ``Living Systematic Review'', a paradigm long advocated for in medical and computing sciences to combat the rapid obsolescence of findings \cite{10.1371/journal.pmed.1001603}. Recent empirical work confirms that the primary barrier to living reviews is the prohibitive cost of updating, with manual re-screening viewed as ``unmanageable'' without technological support \cite{10.1145/3706598.3714047}. \sysname{} alters this cost structure by transforming the review methodology from a static text description into an executable protocol. Because the system stores both the complex, multi-database search strings and the guided AI screening logic (derived from the user's few-shot examples), an update becomes a computational operation rather than a manual reconstruction. A researcher can return to \sysname{} months after publication, re-execute the stored search protocols against new data, and have the AI apply the established screening criteria to \textit{only} the newly published papers. This capability operationalizes the transition from static ``point-in-time'' documents to dynamic knowledge resources that evolve alongside the field \cite{WOHLIN2022106908}.

Finally, we must critically situate these design affordances within the limitations of our current evaluation. Our controlled study demonstrated that users trust and value the system's transparency in short bursts, but the dynamics of reliance in a longitudinal context require careful consideration. As automation becomes more capable, there is a distinct risk of over-reliance on the AI's relevance scoring during long-term maintenance \cite{lee2004trust}. \sysname{} aims to mitigate this through its \textit{Suggestion with Rationale} design, which enforces a moment of verification rather than passive acceptance. Furthermore, while our current single-user evaluation limits our ability to claim full support for large-scale, collaborative team science, a standard process for large SLRs \cite{stokols2008science}, the underlying data structures in \sysname{} are designed to feed directly into collaborative ecosystems (e.g., via standard \texttt{.bib} exports). Future work must expand this infrastructure to support multi-user adjudication, ensuring that the ``living review'' remains a rigorous, human-centered knowledge discovery process.

\subsection{Call for a More Open Academic Infrastructure}
The fragmented nature of the academic ecosystem is a fundamental challenge extending beyond any single tool. The effectiveness of synthesis systems is directly limited by data availability. Our experience building Arc revealed barriers that hinder innovation and contradict Open Science principles. Our experience building \sysname{} revealed various barriers that could potentially hinder innovation and contradict the principles of open science. We argue that the HCI community must advocate for a more accessible and interoperable infrastructure for scholarly knowledge.

We faced two primary obstacles during development. The first is the lack of robust, public APIs for essential academic databases. Key resources for computing research, such as the ACM Digital Library (ACM DL), do not offer a programmatic way to access their well-cataloged academic data in bulk. This forces researchers and tool builders into an undesirable choice. They either omit a critical data source, thereby compromising the tool's comprehensiveness, or rely on customized web scraping techniques (i.e., as described by \citet{10.1145/3392866}). Scraping is an unreliable and inefficient solution. It creates unnecessary server load for the provider and is prone to errors, threatening the validity and reproducibility of the research it is meant to support.

Second, even when APIs are available, access can be prohibitively difficult for academic researchers. Our team spent over six months in a protracted process of emails and follow-ups to acquire an API key for IEEE Xplore. This level of administrative overhead is a significant barrier for many research teams, slowing or preventing the fruition of new scientific discoveries that could benefit the entire community.

To address these challenges, we call on the community and publishing partners to embrace the FAIR Guiding Principles for scientific data, ensuring that it is Findable, Accessible, Interoperable, and Reusable \cite{wilkinson2016fair, dierkes2017fair}. This leads to three concrete recommendations:
\begin{enumerate}
    \item \textbf{Public APIs as a Standard:} Major academic publishers and database providers (e.g., ACM, IEEE) should offer robust, publicly documented APIs for programmatic data access. This should be a standard service for the research community, not a special exception.
    \item \textbf{Simplified Access for Researchers:} A clear and efficient process must be established for academic researchers to obtain API credentials for non-commercial projects. Gaining access should be a matter of days (or with a set turnaround time), not months, enabling small/marginalized teams and independent scholars to innovate without prohibitive unforeseen delays.
    \item \textbf{Improved Interoperability:} Key academic data providers should work toward greater consistency in their data offerings. While full standardization is a long-term goal, progress can be made through the use of common metadata schemas, shared query parameters, and standardized formats. Reducing the need for researchers and developers to handle inconsistent data structures. This will directly support effective user-facing tools and/or accelerate scientific discovery.
\end{enumerate}

Eventually, the friction in the data layer is inevitably transferred to the user interface, and the cognitive load on the researchers. As a community dedicated to designing better human-computer interactions, our work extends beyond the frontend. We must also be advocates for the open, accessible, and interoperable epistemic infrastructure that will enable the next generation of tools for thought.



\section{Limitation and Future Work}

While our findings demonstrate \sysname{}'s potential to improve the SLR workflow, we acknowledge several limitations, which in turn offer clear directions for future research.

First, our comparative user study was limited by its scale and scope. The sample was small ($N=8$), time-constrained, and drew exclusively from computing researchers. Consequently, the results should be read as directional evidence about interaction mechanisms rather than a promise of end-to-end savings, nor have we validated the design's efficacy in domains with differing methodological norms. The study was designed to isolate high-frequency operations, such as query iteration, result consolidation, and initial screening, because these tasks were most commonly performed. However, the real systematic review process introduces additional aspects we did not observe, including multi-reviewer coordination, criteria drift, and downstream data extraction. While our objective metrics for the snowballing and filtering task showed signs of efficiency gains, these were derived from a controlled batch of 15 papers. A longitudinal, in-the-wild deployment is necessary to validate whether these benefits are sustained over thousands of papers and to observe potential new costs. In particular, long-term SLR work raises challenges beyond individual efficiency, such as maintaining auditable decision histories, resolving reviewer disagreement, and supporting the evolution of search strategies and inclusion criteria over time. Future research should examine how \sysname{}'s history, comparison, and rationale-tracking mechanisms function as persistent analytical artifacts in collaborative, months-long review efforts rather than short, isolated tasks.

Second, the system’s performance is strictly bound by the coverage and structure of external data. While \sysname{} orchestrates queries, the absolute search comprehensiveness depends on the selected API endpoints (e.g., Semantic Scholar, IEEE). Therefore, our evaluation focused on interaction utility rather than bibliometric recall, and future research should investigate optimal academic API combinations to maximize coverage against gold-standard SLR datasets. 

Third, the effectiveness of our approach currently relies on prompt engineering tailored to scientific abstracts; future work must evaluate how this transfers to distinct domains, such as law or the humanities. Additionally, we did not explicitly characterize granular failure modes, such as systematic misclassifications arising from ambiguous text or over-generalized reasoning. Crucially, to mitigate the risk of automation bias and over-reliance as trust is established \cite{kim2025fostering}, future research must investigate cognitive forcing functions to structurally enforce the human oversight necessary for failure-aware collaboration.

Finally, we aim to explore novel interactions that advance \sysname{}'s capabilities as a tool for thought. We note that our reported efficiency metrics represent a snapshot of current model capabilities. As LLMs continue to advance in processing speed and reasoning accuracy, we anticipate the time required for AI generation will decrease while suggestion quality improves, likely widening the efficiency gap between AI-assisted and manual workflows. Future work should leverage these advancements to support higher-order cognitive tasks, such as synthesis and hypothesis generation, moving beyond procedural support to become a collaborator in scholarly inquiry.



\section{Conclusion}
Systematic Literature Reviews (SLRs) are foundational to scientific progress, yet the process is often hindered by fragmented workflows that impose high cognitive load. Informed by the insights of 20 experienced researchers through our exploratory design study, we created an integrated design probe to understand how a cohesive, intelligent environment supports this iterative process. Our comparative evaluation suggests that reducing logistical barriers does more than improve efficiency; it enables researchers to shift focus from administrative overhead to strategic decision-making. By facilitating transparent search comparison and verifiable AI-assisted screening, \sysname{} encourages a more experimental and confident approach to literature exploration. These findings suggest that the value of future academic tools lies not just in better search algorithms, but in the interaction design that connects them. As the scientific record continues to outpace our manual capacity, future tools must do more than just manage the scale, they must safeguard the researcher’s ability to find meaning within it -- ensuring that the future of discovery is defined by the creativity and ingenuity of scholars, rather than the time spent fixing spreadsheets.



\begin{acks}
We acknowledge and thank the support of the Natural Sciences and Engineering Research Council of Canada (NSERC), [funding reference number RGPIN-2024-04348 and RGPIN-2024-06005].

Furthermore, we acknowledge and thank Mohi Reza and Ananya Bhattacharjee for their valuable discussion and insights. We extend a heartfelt thanks to our study participants, without whom this work would not have been possible.
\end{acks}


\bibliographystyle{ACM-Reference-Format}





\appendix

\onecolumn


\section{Appendix}

\subsection{Exploratory Design Study Participant Demographic and Library Usage}
\label{appendix:formative-demograpghic}

\input{tables/formative-study-participant-meta}


\begin{figure*}[ht!]
    \centering
    \includegraphics[width=.85\linewidth]{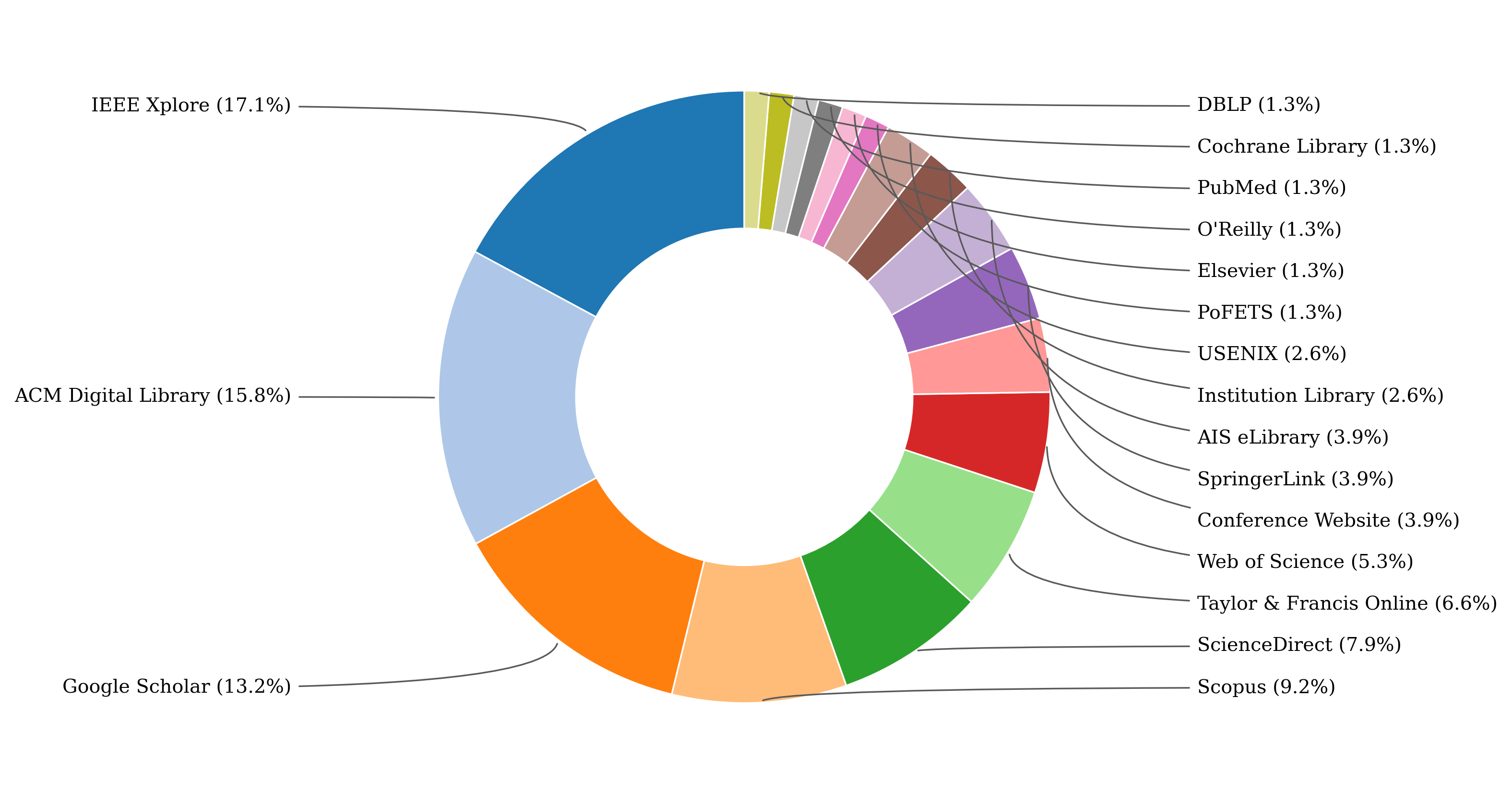}
    \caption{Distribution of Academic Database Usage. This pie chart displays the frequency for different libraries mentioned during the interview, with 17 of the 20 researcher interviewees sharing the database(s) they used during their SLR process.}
    \label{fig:formative-database-usage}
    \Description{This pie chart shows the frequency of usage of different academic databases during the SLR process, as mentioned by 17 researcher interviewees. IEEE Xplore is the most frequently used, followed by ACM Digital Library and Google Scholar, indicating the preferences among the researchers for conducting systematic literature reviews.}
\end{figure*}

\hfill

\newpage
\hfill

\subsection{Prompt Used for LLM-Assisted Paper Filtering}
\label{sec:appendix-prompt-for-llm-filtering}

\begin{lstlisting}[basicstyle=\ttfamily\small,
                   breaklines=true,
                   frame=single,
                   backgroundcolor=\color{gray!10},
                   caption={LLM Filter Prompt Template},
                   captionpos=b,
                   numbers=none]
You are provided with an academic publication's detailed information along with its metadata. Your task is to carefully analyze the data and accurately answer the following questions. Review the provided publication metadata, especially the title and abstract, to determine the best answer for this paper.

Provide detailed justification and rationale for your selected answers, referencing specific parts/phrases of the metadata where appropriate.

Instructions: 
- For each question, select one of the comma-separated possible answers. 
- Ensure that your choice is fully supported by the details in the publication data.

Questions:
{qna}

Examples:
{examples}

Publication Data:
{paper_data}

Output Format:
{format_instructions}
\end{lstlisting}

\newpage

\onecolumn

\subsection{Comparative User Study Workflow}
\label{appendix:user-study-procedure}
\begin{figure*}[ht]
    \centering
    \includegraphics[width=1\linewidth]{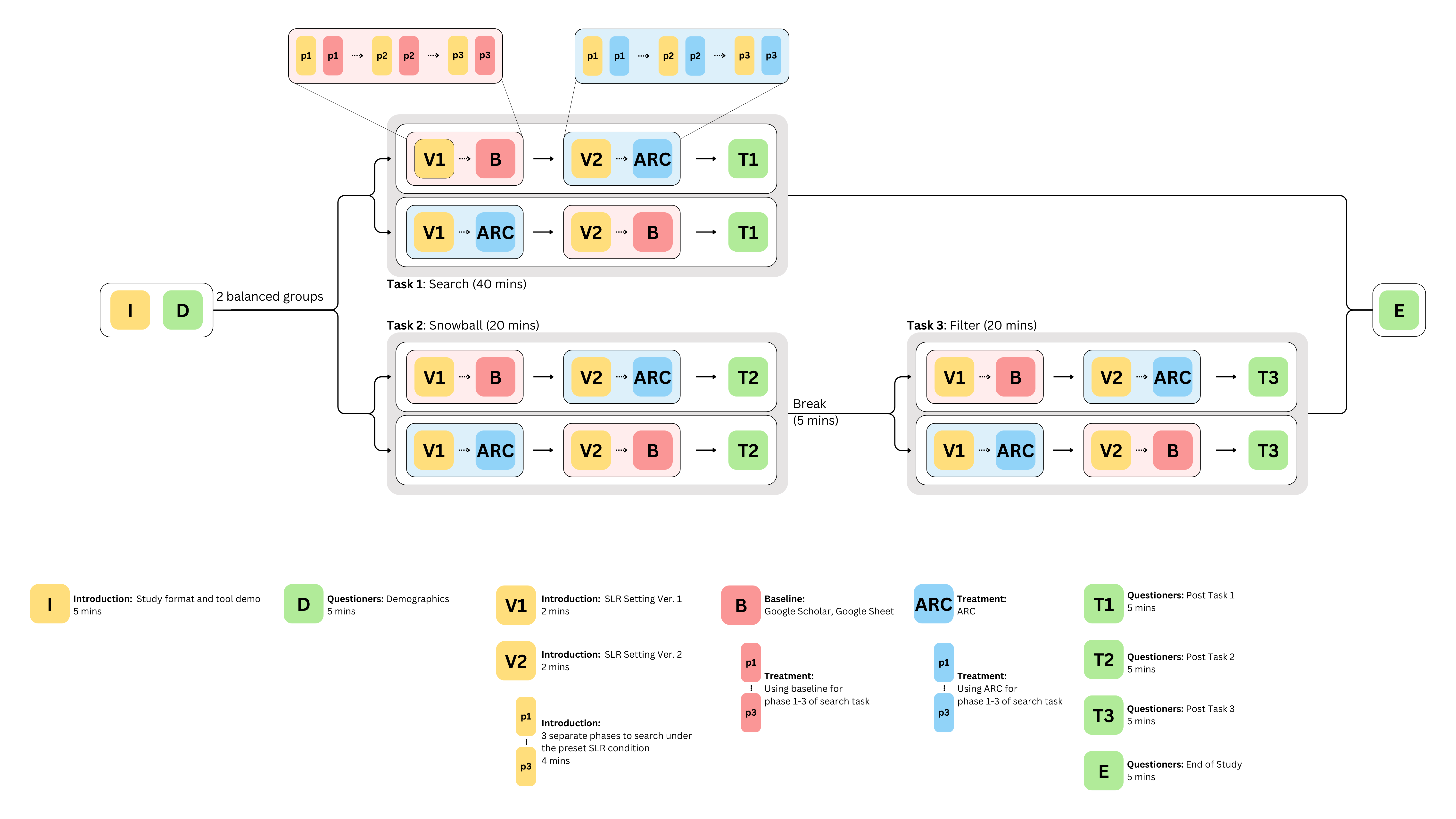}
    \caption{Overview of the Comparative User Study Procedure. After a brief introduction, participants were divided into two balanced groups to tackle different phases of the SLR workflow, structured into three distinct tasks. For each task, participants were introduced to one of two SLR settings, completed the task with either the baseline or \sysname{} condition, then switched to the other setting and condition. Post-task questionnaires were administered following each task.}
    \Description{A flowchart illustrating the comparative user study procedure. Following a brief introduction, participants are divided into two balanced groups. The workflow moves through three distinct tasks: Task 1 (Search), Task 2 (Snowballing), and Task 3 (Filter). For each task, participants alternate between two conditions: using the Baseline tool and using the Arc tool. The order of tool exposure and the specific research scenario are counterbalanced across the groups to ensure a valid comparison. Post-task questionnaires are administered after each block.}
    \label{fig:user-study-procedure}
\end{figure*}

\newpage

\subsection{Comparative User Study Demographic}
\label{appendix:user-eval-demograpghic}


\begin{table*}[ht!]
\centering
\small
\begin{tabular}{@{}llllll@{}}
\toprule
Participant & Gender & Academic Status & Years of Research & Assigned Task & Condition \\
\midrule
P1 & Not Specified & Graduate student (PhD) & 2 & Searching & Baseline $\rightarrow$ \sysname{} \\
P2 & Man & Graduate student (Master’s) & 4 & Searching & \sysname{} $\rightarrow$ Baseline \\
P3 & Woman & Graduate student (PhD) & 9 & Searching & Baseline $\rightarrow$ \sysname{} \\
P4 & Man & Graduate student (PhD) & 4 & Searching & \sysname{} $\rightarrow$ Baseline \\
P5 & Man & Graduate student (PhD) & 2 & Snowballing and Filtering & Baseline $\rightarrow$ \sysname{} \\
P6 & Man & Graduate student (PhD) & 5 & Snowballing and Filtering & \sysname{} $\rightarrow$ Baseline \\
P7 & Man & Graduate student (PhD) & 1 & Snowballing and Filtering & Baseline $\rightarrow$ \sysname{} \\
P8 & Man & Assistant Professor & 12 & Snowballing and Filtering & \sysname{} $\rightarrow$ Baseline \\
\bottomrule
\end{tabular}
\caption{Comparative User Study: Participant Demographics}
\label{tab:demographics}
\end{table*}

\begin{table*}[ht]
\centering
\begin{tabular}{@{}lllll@{}}
\toprule
Participant & General SLR & Advanced Search & Snowballing & Inclusion/Exclusion \\
\midrule
P1 & Strongly agree & Strongly agree & Agree & Strongly agree \\
P2 & Agree & Agree & Agree & Somewhat agree \\
P3 & Somewhat agree & Agree & Agree & Neither agree nor disagree \\
P4 & Strongly agree & Somewhat agree & Somewhat agree & Strongly agree \\
P5 & Somewhat agree & Agree & Somewhat disagree & Agree \\
P6 & Strongly agree & Somewhat disagree & Agree & Agree \\
P7 & Agree & Strongly agree & Agree & Agree \\
P8 & Agree & Agree & Somewhat agree & Somewhat agree \\
\bottomrule
\end{tabular}
\caption{Comparative User Study: Participant Familiarity with SLR}
\label{tab:baseline_familiarity}
\end{table*}

\begin{table*}[ht]
\centering

\begin{tabular}{@{}lll@{}}
\toprule
Participant & Google Scholar & Google Spreadsheets \\
\midrule
P1 & Somewhat agree & Strongly agree \\
P2 & Agree & Agree \\
P3 & Strongly agree & Agree \\
P4 & Agree & Agree \\
P5 & Agree & Somewhat agree \\
P6 & Agree & Somewhat agree \\
P7 & Agree & Strongly agree \\
P8 & Strongly agree & Strongly agree \\
\bottomrule
\end{tabular}
\caption{Comparative User Study: Participant Familiarity with Baseline Research Tools}
\label{tab:demographics_tools}
\end{table*}


\newpage

\subsection{Comparative User Study Statistics}
\label{appendix:user-study-stats}

\input{tables/user-study-stats}

%
\newpage

\subsection{\sysname{} Technical Flowchart}
\begin{figure*}[ht]
    \centering
    \includegraphics[width=\linewidth]{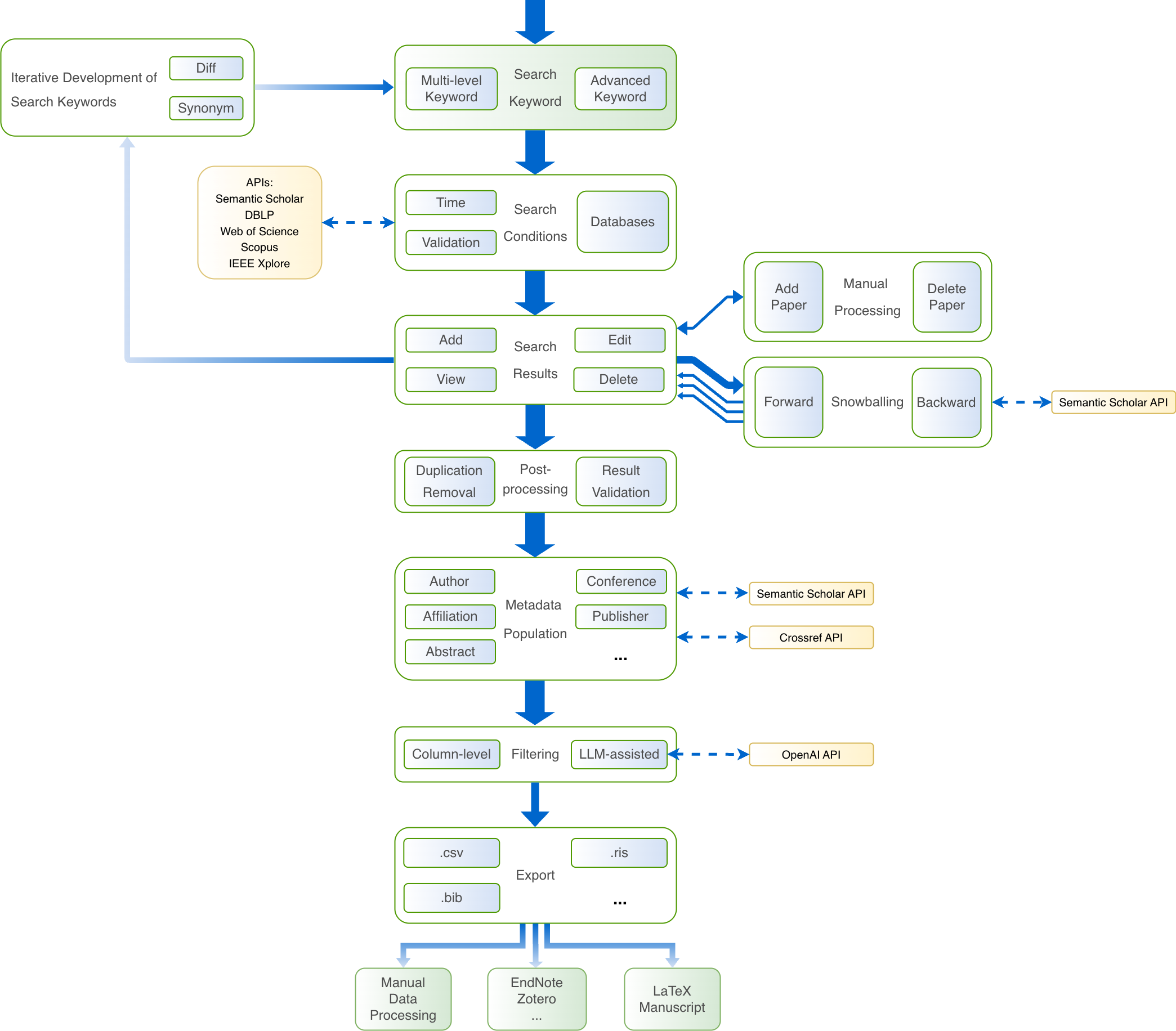}
    \caption{Technical workflow for \sysname{}.}
    \Description{A vertical diagram detailing the technical architecture and workflow of Arc. It starts with the iterative development of search keywords, including diffs and synonyms. This feeds into a multi-level keyword search interacting with various external APIs like Semantic Scholar, DBLP, and IEEE Xplore. The results are processed through validation and de-duplication steps. The workflow supports snowballing via the Semantic Scholar API. Metadata is populated using Crossref and Semantic Scholar. The process concludes with LLM-assisted filtering and data export in formats like CSV, RIS, and BibTeX for use in external reference managers.}
    \label{fig:arc-flowchart}
\end{figure*}


\end{document}

%% file: tables/formative-study-participant-meta.tex
\begin{table*}[ht]
\centering
\begin{tabular}{@{} l l l l p{4cm} l p{3.5cm} @{}}
\toprule
\textbf{\#} & \textbf{Country} & \textbf{Position} & \textbf{Degree} & \textbf{Area of Expertise} & \textbf{Rsrch Exp.} & \textbf{SLR Experience} \\ 
\midrule
R1 & Canada & Asst. Prof. & Master & CER (Applied Machine Learning) & 5--10 yrs & Some experience\\ 
\addlinespace
R2 & Italy & Ph.D. Student & Master & AI \& Cybersecurity (Ethics and Social) & 5--10 yrs & Experienced\\ 
\addlinespace
R3 & Netherlands & Asst. Prof. & Ph.D. & HCI (Databases, Machine Learning) & 5--10 yrs & Some experience \\ 
\addlinespace
R4 & Germany & Postdoc & Ph.D. & DB (Graph DB) & 5--10 yrs & Experienced \\ 
\addlinespace
R5 & Switzerland & Postdoc & Ph.D. & HCI (Privacy) & 5--10 yrs & Highly experienced \\ 
\addlinespace
R6 & Indonesia & Asst. Prof. & Ph.D. & Software Engineering \& CER & 10--20 yrs & Some experience \\ 
\addlinespace
R7 & UK & Researcher & Master & CER & 5--10 yrs & Some experience \\ 
\addlinespace
R8 & USA & Other & Master & Educational Technology & 10--20 yrs & Highly experienced \\ 
\addlinespace
R9 & Canada & Assoc. Prof. & Ph.D. & CER (GenAI in CSEd) & 10--20 yrs & Highly experienced \\ 
\addlinespace
R10 & Finland & Researcher & Ph.D. & CER (Programming Data \& LLM) & 5--10 yrs & Highly experienced \\ 
R11 & Austria & Assoc. Prof. & Ph.D. & Educational Technology \& HCI & 10--20 yrs & Highly experienced \\ 
\addlinespace
E12 & USA & Asst. Prof. & Ph.D. & CER (CS1) & 10--20 yrs & Some experience \\ 
\addlinespace
R13 & India & Ph.D. Student & Other & CER (Autism, Pedagogy, Inclusion) & 3--5 yrs & Highly experienced \\ 
\addlinespace
R14 & Australia & Lecturer & Ph.D. & Formerly Computer Algebra & 10--20 yrs & Experienced \\ 
\addlinespace
R15 & USA & Assoc. Prof. & Ph.D. & CER (AI in Ed, Learning Analytics) & 20--30 yrs & Experienced \\ 
\addlinespace
R16 & Canada & Asst. Prof. & Ph.D. & HCI (data visualization) & 5--10 yrs & Highly experienced \\ 
\addlinespace
R17 & NZ & Assoc. Prof. & Ph.D. & CER (learner sourcing, LLM) & 20--30 yrs & Highly experienced \\ 
\addlinespace
R18 & Brazil & Teaching Prof. & Ph.D. & HCI (Digital Design \& Tangible Interface) & 10--20 yrs & Experienced \\ 
\addlinespace
R19 & Canada & Postdoc & Ph.D. & HCI (AR, VR, Eye tracking) & 5--10 yrs & Some experience\\ 
\addlinespace
R20 & Ecuador & Asst. Prof. & Ph.D. & ML (Optimization) & 5--10 yrs & Highly experienced \\ 
\bottomrule
\end{tabular}
\Description{This table provides demographic data for researcher interviewees, categorized by their systematic review experience. The experience is divided into three categories: (1) Some experience (little or no formal systematic review experience), (2) Experienced (at least one formal peer-reviewed systematic review publication), and (3) Highly experienced (two or more formal peer-reviewed publications). The table also lists the participants' research experience (in years), areas of expertise (ranging from machine learning to educational technology), current positions, highest degrees, and countries of residence.}
\caption{Formative Study Researcher Interviewee Demographic. We categorize their systematic review experience into three categories: (1) Some experience: where the interviewee did not have any experience conducting formal SLR, is currently working on one, and/or has other informal literature review experience; (2) Experienced: where the interviewee has one formal peer-reviewed SLR publication; (3) Highly experienced: where interviewee have two or more formal peer-reviewed SLR publications.}
\label{tab:interview-metadata}
\end{table*}

%% file: tables/user-study-stats.tex
\begin{table*}[ht]
\centering
\label{tab:survey_stats_taps}
\begin{tabular}{p{7cm} l c c c}
\toprule
\textbf{Question} & \textbf{System} & \textbf{Mean} & \textbf{SD} & \textbf{95\% CI} \\ 
\midrule

\multirow{2}{7cm}{\textbf{Q1:} Complete SLR tasks quicker than expected} 
& Baseline & 3.12 & 1.55 & 1.30 \\
& \sysname{} & \textbf{5.75} & 1.04 & 0.87 \\ 
\midrule

\multirow{2}{7cm}{\textbf{Q2:} Took less effort than usual} 
& Baseline & 3.00 & 1.31 & 1.09 \\
& \sysname{} & \textbf{5.62} & 1.60 & 1.34 \\ 
\midrule

\multirow{2}{7cm}{\textbf{Q3:} Interface was straightforward and intuitive} 
& Baseline & 4.50 & 1.77 & 1.48 \\
& \sysname{} & \textbf{5.75} & 0.71 & 0.59 \\ 
\midrule

\multirow{2}{7cm}{\textbf{Q4:} Did not need strong technical skills} 
& Baseline & 4.50 & 1.31 & 1.09 \\
& \sysname{} & \textbf{5.75} & 0.71 & 0.59 \\ 
\midrule

\multirow{2}{7cm}{\textbf{Q5:} Layout and presentation helped me stay organized} 
& Baseline & 3.38 & 1.92 & 1.61 \\
& \sysname{} & \textbf{5.75} & 1.58 & 1.32 \\ 
\midrule

\multirow{2}{7cm}{\textbf{Q6:} Provided clear guidance for my research workflow} 
& Baseline & 2.62 & 1.06 & 0.89 \\
& \sysname{} & \textbf{5.00} & 1.07 & 0.89 \\ 
\midrule

\multirow{2}{7cm}{\textbf{Q7:} Easy to track and revisit my progress} 
& Baseline & 1.88 & 1.13 & 0.94 \\
& \sysname{} & \textbf{6.12} & 0.99 & 0.83 \\ 
\midrule

\multirow{2}{7cm}{\textbf{Q8:} Perform consistently across different research topics} 
& Baseline & 5.38 & 1.51 & 1.26 \\
& \sysname{} & \textbf{5.88} & 0.64 & 0.54 \\ 
\midrule

\multirow{2}{7cm}{\textbf{Q9:} I felt comfortable relying on the results} 
& Baseline & 4.75 & 1.83 & 1.53 \\
& \sysname{} & 4.75 & 1.39 & 1.16 \\ 
\midrule

\multirow{2}{7cm}{\textbf{Q10:} Using the tool made the SLR process less stressful} 
& Baseline & 3.25 & 0.71 & 0.59 \\
& \sysname{} & \textbf{6.00} & 0.53 & 0.45 \\ 
\midrule

\multirow{2}{7cm}{\textbf{Q11:} I would like to use the tool again for future SLR} 
& Baseline & 3.00 & 1.69 & 1.41 \\
& \sysname{} & \textbf{6.12} & 0.83 & 0.70 \\ 
\midrule

\multirow{2}{7cm}{\textbf{Q12:} Overall, I am satisfied with my experience} 
& Baseline & 3.12 & 1.13 & 0.94 \\
& \sysname{} & \textbf{6.25} & 0.89 & 0.74 \\ 
\bottomrule
\end{tabular}
\caption{Comparison of User Survey Responses (1-7 Scale) between Baseline and \sysname{}. Statistics include Mean, Standard Deviation (SD), and 95\% Confidence Interval (CI).}
\end{table*}